\documentclass{article}
\usepackage[a4paper,margin=1in]{geometry}
\usepackage{url}
\usepackage[square,numbers]{natbib}

\usepackage{anyfontsize} 
\usepackage{enumitem}
\usepackage{mdframed}
\usepackage{tcolorbox}
\definecolor{myabstract}{cmyk}{.04,.04,.12,.08}
\newtcolorbox{myabsbox}{
    arc=0pt,
    boxrule=0pt,
    colback=myabstract,
    width=\textwidth,   
    colupper=black,
}
\usepackage{url}
\makeatletter
\def\url@leostyle{%
  \@ifundefined{selectfont}{\def\UrlFont{\sf}}{\def\UrlFont{\footnotesize\sffamily}}}
\makeatother
\usepackage{xcolor} 
\definecolor{darkgreen}{rgb}{0,0.5,0}
\definecolor{darkred}{rgb}{0.7,0,0}
\usepackage[colorlinks=true,
	urlcolor=blue,
	citecolor=darkgreen,
	linkcolor=darkred,
	pdfencoding=auto,psdextra
]{hyperref}

\usepackage{doi}

\usepackage{tikz}

\usetikzlibrary {backgrounds}
\usetikzlibrary {arrows.meta}
\usetikzlibrary {calc}
\usetikzlibrary {decorations.markings,decorations.pathmorphing}
\usetikzlibrary {graphs}
\usetikzlibrary {intersections}
\usetikzlibrary {math}
\usetikzlibrary {positioning}
\usetikzlibrary {shapes}

\newcommand{\markSB}[1]{%
  \vcenter{\hbox{%
    \tikz[inner sep=2pt]{
      \node (X) {$#1$};
      \draw (X.north west) -- (X.north east);
      \draw (X.north east) -- (X.south east);
    }%
  }}%
}

\newcommand{\markVK}[1]{%
  \vcenter{\hbox{%
    \tikz[inner sep=2.5pt]{
      \node (X) {$#1$};
      \draw (X.north west) -- (X.north east);
      \draw (X.north east) -- (X.south east);
      \draw (X.south east) -- (X.south west);
      \draw (X.south west) -- (X.west);
    }%
  }}%
}

\begin{document}

\title{Open Questions about Time and Self-reference in Living Systems}

\author{
Samson Abramsky$^{1}$,
Wolfgang Banzhaf$^{2,3}$,
Leo S. D. Caves$^{4}$,
Michael Levin$^{5}$,\\[4pt]
Penousal Machado$^{6}$,
Charles Ofria$^{2,3}$,
Susan~Stepney$^{7}$,
Roger White$^{8}$\\[4pt]
\footnotesize
$^{1}$ Department of Computer Science, University College London, 
WC1E 6BT, 
UK\\\footnotesize
$^{2}$ Department of Computer Science and Engineering, Michigan State University, East Lansing, MI 48864, USA\\\footnotesize
$^{3}$BEACON Center
for the Study of Evolution in Action and Ecology, Evolution and Behavior Program, \\\footnotesize
Michigan State University, East Lansing, MI 48864, USA\\\footnotesize
$^{4}$ Independent Researcher, Porto, Portugal\\\footnotesize
$^{5}$ Allen Discovery Center, and Department of Biology, Tufts University, Medford, MA, USA\\\footnotesize
$^{6}$ CISUC/LASI, DEI, University of Coimbra, Coimbra, Portugal\\\footnotesize
$^{7}$ Department of Computer Science, University of York, 
YO10 5DD, 
UK\\\footnotesize
$^{8}$ Department of Geography, Memorial University of Newfoundland, St. John's, NL, Canada
}

\date{}
\maketitle


\begin{abstract}
Living systems exhibit a range of fundamental characteristics: they are active, self-referential, self-modifying systems.
This paper explores how these characteristics create challenges for conventional scientific approaches and why they require new theoretical and formal frameworks. 
We introduce a distinction between `natural time', the continuing present of physical processes, and `representational time', with its framework of  past, present and future that emerges with life itself. Representational time enables memory, learning and prediction, functions of living systems 
essential for their survival.  
Through examples from evolution, embryogenesis and metamorphosis we show how living systems navigate the apparent contradictions arising from self-reference as natural time unwinds self-referential loops into developmental spirals. 
Conventional mathematical and computational formalisms struggle to model self-referential and self-modifying systems
without running into paradox.
We identify promising new directions for modelling self-referential systems, including domain theory, co-algebra, genetic programming, and self-modifying algorithms. There are broad implications for biology, cognitive science and social sciences, because self-reference and self-modification are not problems to be avoided but core features of living systems that must be modelled to understand life's open-ended creativity.

\end{abstract}

\section{Introduction and Motivation}

\label{sec:introduction}

\begin{quote}
    ``Perhaps the first lesson to be learned from biology is that there are lessons to be learned from biology'' 
    
    \hfill --- Robert Rosen, Essays on Life Itself \cite[p.275]{RosenEssays}
\end{quote}

Since the rise of modern science and its success in explaining natural phenomena, there have been two contrasting views regarding its potential domain of application.  On the one hand is the belief that all phenomena, including those characterising biological and social systems, 
can in principle be understood using the approach of conventional science; 
the most extreme position is that every legitimate explanation must
ultimately be reducible to fundamental physical laws.   
On the other hand is the belief that biological and social phenomena, although themselves physical in basis, exist in such {a Vanishingly Small and atypical fraction of the overall Vast\footnote{%
Dennett \cite{dennett1996} uses the capitalisation of the terms \textit{Vast} and \textit{Vanishingly Small} to constantly remind the reader of the almost unimaginable sheer combinatorial hugeness of evolutionary design space, and the resulting still absolutely huge but nevertheless relatively tiny subset of that space that is explorable.
}} state space of physical material and processes
that they need their own additional and different explanatory mechanisms.

{The overwhelming majority of possible molecules lack stable structure or meaningful functional activity, limiting living systems to an atypical subset where replication is possible.
 The complexity and rarity of self-replication makes the origins of life difficult to explain, and constrains the pathways available to evolution.  Identifying which other properties that are disproportionately common in these atypical regions will help us understand how evolutionary processes unfold over time
\cite{Higgs2015,Sole-2025}.
Life can undergo major evolutionary transitions \cite{MaynardSmith1995,Szathmary2015} where formerly distinct living individuals unite together into higher-level organisms.  Two kinds of prokaryotic cells merged to form eukaryotic cells (where one became the mitochondria).  Eukaryotic cells evolved to `stay together' and became multicellular organisms.  These major transitions represent new kinds of individuals that, in most cases, are far more complex than their ancestors and have yet more distinct properties in their behavior and evolution.  In this sense, major transitions can be understood as movements into increasingly constrained and highly structured regions of possibility space, where new kinds of processes become viable.}

{Our focus on time, self-reference, and representational dynamics complements this framework by emphasising the processes through which such transitions may be realised and sustained.}
We develop our position by exploring fundamental characteristics of living systems that set them apart from typical physical systems and thus need both an extension of the conventional scientific paradigm and the development of new tools to investigate the consequences of these differences.  
Here we reserve the term  \textit{physical systems} to refer to those systems in the typical region of the possibility state space, 
and use the term {\it living systems} to refer to both biological and social systems, including the emergent phenomena of social systems such as the arts, technology, and science itself.

\subsection{Philosophical stances}
There are many philosophical stances one can take when discussing the properties and processes of life.
Some of these stances relevant to this paper include:
\textit{representationalism} \cite{Lycan2023},
that sensory experience of the world
is mediated through internal representations or models of the world;
{the \textit{free-energy principle} \cite{Friston2010} and}
\textit{predictive coding theory} \cite{Clark2013},
that brains are `predictive machines' that attempt to minimise {surprise or} error between external sensory signals and internal sensory predictions, arguably a form of representationalism \cite{Gladziejewski2016};
\textit{embodiment} \cite{Pfeifer2006,SS-Embody},
that living entities are closely coupled with or embodied in their environment 
{(encompassing both brain-body and body-external environment couplings)},
in a complex feedback relationship,
and through this coupling are both constrained and enabled by their environment shaping their own behaviours;
and \textit{enactivism} \cite{Varela1991b},
that the entity and its environment fully co-create each other through reciprocal coupling.
These stances are typically taken in the context of studying consciousness and cognition,
but they can be applied at all levels of life, from single cells up.
For example, it is common to consider that the DNA of an organism encodes some form of information that is a representation of, correlated with, its environment.

These various stances have one thing in common relevant here:
they all involve a form of dynamic coupling of self and other:
coupling via an internal model of the other, via physical embodiment, via deeply connected co-creation.
This coupling, however described, is essential to, and deeply shapes, the living entity.

When one takes a particular philosophical stance to talk about things, 
one is talking about them \textit{as if} they have the properties relevant to that stance, without them necessarily having those properties.
For example, one can take an intentional stance  \cite{Dennett1987} 
and describe a thermostat as intending to keep the room temperature constant,
without needing to believe or require the thermostat has that intention, or any intention at all.
Taking a particular stance is an explanatory strategy.

{In this paper, we take diverse philosophical stances, depending on the point we wish to make, and on each particular coauthor's own position.}
We typically take a representational stance, but sometimes  an embodiment stance, and sometimes an enactivist stance.
No one stance tells the whole story;
each is an abstraction, an incomplete facet;
together they may provide a fuller picture of actual living systems.

\subsection{Living systems, self-reference, and time}

Living systems differ from typical physical systems in that they are active self-referential, self-modifying systems.

A representational stance might explain living systems' self-referential properties in terms of having some form of \textit{representation}, {some internal model,} of themselves and their environment, which they exploit for their existence and behaviour. 
Once a living system's representation of itself includes its representation of itself, it is \textit{self-referential}.  
A more enactivist stance might explain living systems' self-reference through organisational closure or circular causality, 
where the system operations result in
maintenance and modification of the organization that produces those operations.

Self-reference can support rules that directly or indirectly govern the behaviour of the system (explained either as represented rules, or as emergent rules, depending on stance)
in a manner analogous to the laws of physics.
Unlike the laws of physics, however, these rules are neither inviolable nor unchanging\footnote{Whether the laws of physics are unchanging can be subject to a lively debate \cite{webb2001,uzan2011}. For the discussion here, we mean {\it unchanging} on a time-scale accessible to humans.}.
Living systems can use self-reference to \textit{self-modify}.  Complex organisms develop from a single cell; over multiple generations organisms evolve to become different ones, with altered DNA. Similarly, societies undergo continual changes in social structures, develop new types of organizations, and invent new technologies which in turn may alter their social structure and organizations.  
The self-modifications that living systems undergo can result in changes to 
the rules that govern their behaviour. 

As Rosen \cite{rosen1991life} points out, this self-reference is 
of the kind that is considered pathological in conventional logic because it can lead to paradox.  
However, the apparent paradoxes introduced by self-reference
can be solved by explicitly including \textit{time}
in the explanation of these inherently dynamical systems, because time unfolds paradox into a process. 
We take up questions concerning the nature of time 
as they arise in attempts to understand the nature of living systems.  Living systems are characterised most fundamentally by the fact that they generate novelty.  They are not just self-organising systems but systems that transform themselves into other systems, through evolution and other processes, and these transformations arise in time. 

\subsection{Explanatory tools}

Our focus here on self-reference, self-modification and time  extends to a discussion of techniques and methodologies that are required to adequately model, analyse, and understand living systems. 
Some of these techniques are old but not widely adopted, and some are new and yet to be fully explored.
Algorithmic approaches are currently the most widely used of these, because they are available and, unlike conventional mathematics, are inherently temporal.  But much needs to be done to extend the range of phenomena that can be modelled algorithmically. This is especially true in the case of self-modifying algorithms, where it would be useful, for example, to extend the depth of self-modification, 
in order to model not just systems that change their rules, but systems that change the rules by which their rules can be changed, thus enabling the emergence of transformational creativity and novelty \cite{Boden-1990,Banzhaf-2016,Stepney:2021:oee4}.  
Other paths toward this modelling include recently developed mathematical approaches like coalgebra.

Much of our discussion involves biological systems.  
The science of biology is undoubtedly advanced and sophisticated, 
but we suggest it may not be asking all the right questions,
somewhat neglecting those related to self-referential, self-modifying systems.
Authors such as Rosen \cite{rosen1991life,RosenEssays,rosen2000} and Varela \cite{varela1975calculus,varela1979,maturana1991,varela1991} are not mainstream. 
The social sciences are also in need of their own methods to address  self-referential and self-modifying social and technological systems.  Social systems, and their emergent phenomena like technology, evolve very rapidly, and yet we lack formal understanding of their evolution as self-referential, self-modifying systems.   
Again, there are authors working in and adjacent to the field
(for example, Luhmann's autopoietic social systems theory \cite{Luhmann1986}; 
Axelrod and others in computational social science \cite{Gilbert2010};
Watts \cite{Watts2004} and Barabási \cite{Barabasi2016} in social network science).
But still, we need the deeper understanding that these new approaches can offer us.

\subsection{History and structure of the paper}

This paper is the result of a workshop on \textit{Time, Life and Self-reference}, held at Memorial University of Newfoundland in July 2023. The discussions summarized here attempt to make several key contributions to our understanding of living systems, from  different philosophical stances
{as held by the different coauthors}. 
We argue for a new theoretical framework that integrates the concepts of time, self-reference, and self-modification, demonstrating how these are fundamentally interlinked in all living systems, from cells to societies. 
We introduce the crucial distinction between what we call \textit{natural time} and \textit{representational time}, showing how the latter emerges with life itself and enables the anticipatory behaviour characteristic of living systems. 
We identify specific limitations in mainstream mathematical and computational approaches for modelling self-modifying systems, particularly in their ability to handle systems that can change their own rules. 
To help address these issues, we propose new directions for developing formal tools capable of modelling such systems, which are of particular relevance to understanding rapid evolution in social systems and technology. 

The paper proceeds as follows. 
Section \ref{sec:nat-and-rep} examines the fundamental relationship between time and living systems, introducing the crucial distinction between natural and representational time.
Section \ref{sec:life} demonstrates why life is fundamentally temporal, analysing this through the lenses of evolution and embryogenesis. 
Section~\ref{sec:self} explores how living selves come to be defined, how then self-reference emerges, and how self-reference with natural time resolves certain paradoxes.
Section \ref{sec:TLR} explores the deep connections between time, life, and self-reference, particularly focusing on morphogenesis and the maintenance of self through radical change. 
Section \ref{sec:tools} examines existing and needed tools for modelling self-referential systems, including mathematical formalisms and computational approaches. Finally, Section \ref{sec:conclusions} concludes by synthesizing these perspectives and identifying key challenges and opportunities for future research in understanding self-modifying open systems.

\section{Time}
\label{sec:nat-and-rep}
\begin{quote}
    ``What, then, is time? If no one asks me, I know what it is. If I wish to explain it to him who asks me, I do not know.'' 
    
    \hfill --- St. Augustine of Hippo, Confessions, Book 11, Chapter XIV (c.397 CE)\footnote{%
    source: \url{https://sourcebooks.fordham.edu/basis/confessions-bod.asp}
    }
\end{quote}
\noindent
Much ink has been spilt by philosophers, physicists, sociologists, narratologists, and novelists, among others, on the mysteries of time;
they cover topics as diverse as cosmology, general relativity, thermodynamics; absolute time, relative time, biological time, sociological time, narrative time, subjective time; past memory, future predictions, durations and events; measuring time, modelling time;
and even whether time exists at all.
See, for example, \cite{Bergson-1889,Proust-recherche-1913,Hawking-BHoT-1988,Barbour-time-1999,Mermin-GR-2005,Carroll-time-2010,Muller-time-2015,Muller-now-2016,Caves-ch19-2018,Longo2021}.
In this work, we focus on two specific notions, which we call natural time and representational time. 
{These provide a foundation for new approaches to understanding living systems, as well as a justification for new, unconventional methodologies.  Furthermore, they allow us} to disentangle certain  {paradoxes that can arise from modelling self-reference (see section~\ref{sec:self})}.

\subsection{Natural time versus representational time}
\label{subsec:two-notions}

Paradoxes of self-reference arise,
we believe, because the word `time' is used to refer to two quite distinct phenomena that are assumed to be the same. 
The first we call {\it natural time}. This is the time of the physical world in which we and everything else exist in a continually evolving present \cite{Muller-now-2016}. 
{It is the time we experience subjectively, the time St. Augustine could not explain.}
We are unable to travel to the past or skip ahead into the future. 
However, living systems must have some capacity for memory, learning, and prediction in order to survive, and these abilities depend on some form of access to the past and future, as well as the present. 
These concepts have no meaning in the context of natural time, where past and future do not exist independently. 

Enactivist stances allow access to the future through several approaches. 
They can understand \textit{anticipatory} behaviour (action taken in response to a predicted future event) through concepts including 
dynamical coupling with environmental rhythms, 
and temporal niche construction where the anticipation is distributed across organism-environment temporal coupling rather than being localised in internal models.

Here, in contrast, we take a representational stance,
and say that past and future exist as properties of another kind of time,
which we call {\it representational time} \cite{white2020}. 
This representational time appears along with the origin of life, as a necessary property of living systems\footnote{Indeed, many {living sub-systems} also seem to exploit it, for example, simple gene-regulatory networks. 
More generally, many kinds of chemical pathways can exhibit six different types of learning \cite{pigozzi2025}. This does not require being a cell or having neurons.
In this view, representational time might be a non-binary or gradual concept;  it might gradually switch on, or emerge, as systems become (more) self-referential, more alive.
}. 
We stress that, despite its name, representational time is  {more than} simply a representation of natural time: it is a new kind of time, one that permits a view of the world that transcends what is possible with natural time, permitting agents to form memories of the past, make predictions of possible futures, and act on these, rather than merely react to the immediate. It augments natural time, and thus permits consequences in the physical world that could not occur in its absence; in that sense it is `real'.

\newpage
\subsection{Life and representational time}\label{subsec:origin}
\begin{quote}
    \textbf{Is there a new concept of time that comes into being with the origin/development of life?}
\end{quote}

\noindent
In conventional logic and mathematics, self-reference is approached with caution because it can lead to paradox, particularly when the self-reference applies to the whole system making the reference rather than to just a part of the system.
Whole-to-part self-reference is not problematic, but whole-to-whole self-reference may be, though not necessarily.
For example, Russell's paradox \cite{russell2016} arises from the statement ``the set of all sets that do not belong to themselves belongs to itself'', as the set both must be and must not be part of itself. 
Conversely, the claim ``the set of all sets that belong to themselves belongs to itself'' does not create a contradiction and is thus an example of how whole-to-whole self-reference is not inherently problematic.
{Conventional mathematics attempts to solve paradoxes that arise from self-reference simply by banning self-reference.
Since life is fundamentally self-referential,
such a solution is not acceptable in the approaches taken to model it. See section~\ref{sec:tools} for a discussion of other mathematical approaches to self-reference.}

When considering the nature of time, the act of proving of a statement, in the context of the relevant axioms and proof rules, must unfold over natural time, as does any process, but the proof's conclusion, once established, is an immutable truth in that context; it is  timeless, i.e. its continuing validity does not involve natural time. 
In non-problematic cases of self-reference, the proof process leads in a finite number of steps to a definite result, e.g. `true' or `false'. It was the process of proof that Turing was trying to formalise with his concept of the universal computer or Turing machine \cite{yanofsky2013}. But his project failed in the sense that he showed that it could not be proved in general whether an algorithm would terminate in a finite period of time, and thus the possibility remained open that some theorems, while true, could not be proven to be so. The computation process of a Turing machine is constructed of known logic and mathematics, and the failure to halt, i.e. the failure to reach a definite conclusion, may be taken as a sign that additional, unknown, mathematics or logic is required, a result analogous to Gödel's incompleteness theorems in mathematics \cite{yanofsky2013}. But where would this additional mathematics come from? Historically, it has come from the creative imagination of mathematicians.

Mathematicians are living organisms, and living entities are inherently self-referential, from the most basic level of the DNA that encodes information of the organism itself, to the ``I am'' statements of a mathematician. As we see in the biological evolution that has led from single cell organisms that maintain and reproduce themselves to highly complex multi-cellular organisms that do the same, but also do mathematics and write algorithms for Turing machines, both individual organisms and life as a whole are creative.

The mathematical biologist Robert Rosen, seeking to understand the nature of life, recognized that self-reference, or self-modelling, was a fundamental characteristic of life \cite{rahwan2019machine}, but this raised difficulties he could not resolve within the formalisms he considered permissible: those allowed by conventional mathematics frequently led to paradoxes or infinite regresses when applied to the problems of living systems. He concluded that ``life is not an algorithm,'' by which he meant that it could not be modelled by a Turing machine. But Turing machines are constructed from conventional logic and  mathematics; they necessarily produce timeless results, or indeed may produce no result at all.

Limiting permissible formal tools to classical Turing machines eliminates natural time from the results, so that they are stable and certain. But in eliminating natural time we lose the ability to address certain problems formally.  Specifically, we cannot handle problems that involve creative processes, i.e. those for which there is no final result, only a process that continually generates a sequence of novel structures, each one leading to the next, whether these are molecules, life forms, or ideas.

Rosen understood living organisms as {\it anticipatory systems}, i.e. systems with dynamics that are forced by a future state of the system \cite{rosen2000}. Of course, in a literal sense, this is not physically possible. The forcing is actually by an {\it anticipated} future state, where the anticipation is generated by an internal model of the system itself, i.e. by a self-referential model.  However, the model, while useful, can never be entirely correct, because it is a model, so it will occasionally generate errors, and these errors will occasionally function as a source of novelty.

While the processes of the living system, and in particular of the self-referential model, must unfold in natural time, the anticipation generated by the model also requires a representation of time. To the extent that life is anticipatory, it requires a {\it representation} of natural time, and so must itself generate such a representation. Thus representational time is a new kind of time, one which co-originates with life. Unlike natural time, which has only a present, representational time also has a past and a future.

The addition of past and future permits {\it memory} (past events can be stored {and recalled}), {\it anticipation} (the model can predict future events based on past and current events), and {\it learning} (anticipations can be stored in memory to be compared with memories of actual events, and discrepancies between the two can then be used to modify the model that produced the anticipations).

Natural time, which might also be called physical time or the time of the physical universe, consists of a continuing present. Muller conceives of it as the continuing creation and transformation of the universe in the big bang \cite{Muller-now-2016}. This process is continually instantaneous, occurring in a continual present. In physics, this characteristic appears in fundamental expressions like the Hamiltonian, which describes the instantaneous evolution of a physical system. The Hamiltonian is composed of partial derivatives with respect to space and time, and so, with the time interval being infinitesimal, the Hamiltonian is essentially timeless; in that sense it represents the physical universe in natural time.\footnote{%
While the Lagrangian formulation lacks this instantaneous quality \cite{Stepney-IJGS14}, it is nevertheless unable to characterize complex dynamical systems, especially those undergoing open-ended evolution.}

However, the Hamiltonian tells us nothing about what we will observe until we integrate it, so that we can calculate past and future states. This mathematical process of integration introduces an explicit time, a representational time with past, present and future. This points to a fundamental distinction between non-living and living systems. 
All physical systems evolve without direct access to past or future states, only in response to their current state as described by the Hamiltonian. 
Of course, the current state is always a result of past states, but in typical physical systems the evolution into the future requires no \textit{representation} of those past states. Living systems, on the other hand, exploit representations of past states, as well as predictions of future states. Organisms must survive to reproduce, proliferate, and ultimately evolve. This requires the exercise of agency in their environment: they must be able to find food, identify it when they encounter it, and avoid toxins and predators. To do so they must be able to learn and to make predictions. In other words they must have a functional representation of past and future.

Even simple life forms like unicellular organisms contain a number of representations of time.  For relations with the organism's environment, time is represented implicitly by molecular structures that recognize relevant  conditions such as the presence of food or toxins, or chemical or light gradients that it may be beneficial to follow. 
These are learned responses to environmental conditions, but for the most part the learning has occurred during the evolution of the species, so the memories and the anticipatory responses are fixed in the individual; time is represented in the functionality of the organism. 

More fundamentally, the organism represents its own internal processes,  like metabolic activity or reproduction, and these processes require their own representations of time.  Execution of chemical reactions in specific sequences 
means that there must be procedures for timing in a relative, if not an absolute, sense. The internal structure of the cell is designed, again by evolution, to constrain the dynamics of the reaction network so that reactions occur in the proper sequence; enzymes, which modify reaction times, play a key role in this control process. 
More generally, enzyme kinetics provides a means to manipulate reaction times between substrates, and thus allows the introduction of a nuanced timing to the dynamics of the system. 
These time-sensitive internal mechanisms are essential for survival, and they embody a form of temporal representation shaped by evolutionary history.
In complex organisms with nervous systems, dedicated timing mechanisms appear, and establish, for example, diurnal rhythms and, ultimately, representations like relativistic space-time.

Human societies have produced many explicit representations of time; most, if not all, of them include past, present and future. Many domains, but notably science, use a linear representation of time, with time denoted by a variable $t$. This is a powerful representation, because if we are able to express some quantity of interest as a function of $t$, we can immediately solve to find the value at any past or future time, and this in turn enables us to make testable predictions, and the ability to predict is one of the foundations of science. The very power of this representation, however, is also its limitation. As Bergson \cite{bergson1911} points out, the certainty inherent in mathematics leaves no room for novel phenomena to appear.

In other words, if we were to rely purely on the explicit representation of time, it would rule out the possibility of a scientific understanding of strong emergence or open ended evolution. This is the reason that we need new approaches to understand and work with novelty.  Some of these are already emerging, as we discuss in Section~\ref{sec:tools}.

\newpage
\section{Life is essentially temporal}
\label{sec:life}

Although life uses representational time to provide it with memory and predictive powers,
it has many iterative properties that give it temporality and embed it in natural time.

Life is far-from-equilibrium, it is highly structured, both spatially and temporally.  
As noted before, life exists in an extremely atypical (Vanishingly Small probability) region of the entire Vast possibility state space of physical matter and its processes.  What processes might guide a system to find and exploit such an atypical region?
Here we discuss two naturally occurring processes that can give rise to finding this complex place.  {\it Evolution}, acting over timescales of multiple generations, and {\it ontogeny} (growth and development), acting over an individual organism’s lifespan.  These processes appear to be essentially iterative, and hence embodied in natural time.

\subsection{Evolution as iterative generate and test}
\label{subsec:gen-and-test}

At its simplest, Darwinian evolution is expressed as inheritance with variation undergoing natural selection, iterated over multiple generations.  These processes are complex in the living world; for example, variation might be produced by single base mutation of DNA, larger scale changes through transposons, horizontal gene transfer, epigenetic mechanisms, and more.  These processes are themselves embodied and subject to evolution.  
However, at root, Darwinian evolution can be considered an instance of an abstract \textit{generate and test} process\footnote{\label{foot:no-separateion}%
While we can
analytically distinguish processes from the content on which they operate, in
living systems these are not separate entities but different aspects of the
same embodied organisation. The `algorithms' of inheritance, variation, and
selection are themselves materially instantiated in the same biological
structures that constitute the evolving population.
}.  
There is a population of \textit{candidates}: new candidates are {\it generated} from this population, and their suitability is then {\it tested}.  Candidates that pass the tests are more likely to remain in the population, to become sources for the subsequent generation in their turn.
Unlike random search, evolutionary processes retain and build upon successful solutions, as such they have \textit{memory}:
the current candidate population has implicitly become `better' at {passing the test} than previous populations, at least in an unchanging environment.  In natural populations, this enhancement {can be} in the form of either improved survival or reproduction\footnote{{Natural evolution also includes more complex processes, such as neutral drift, and is highly stochastic: passing the `test' includes a large amount of luck.  However, the underlying iterative generate and test structure is present.}}.  This entire process is most effective when the improved population is also associated with additional opportunities for continued progress.

Generate and test is thus essentially an {\it iterative} scheme. At each step the population has changed, and the new population becomes the source of the next generation.  In more self-referential systems, the processes leading to these changes may themselves evolve.  In the most trivial cases, it might be possible to `short-cut' the iterations, directly calculating the final (or at least some far future) result by composing multiple generate and test functions.  However, in cases where the steps are non-trivial, and particularly when they themselves are changed by  previous steps (evolution of the algorithms as well as of its population), the whole process is {\it computationally irreducible}
{in the sense that there is no general procedure for obtaining the state at step $n$ without simulating the steps that produce it}\footnote{%
{In a process with a fixed update rule $f$, the state at step $n$ can sometimes be obtained directly from a closed-form expression for $f$ applied $n$ times.
In a self-modifying process, no such fixed rule exists: the rule applied at step $k+1$ is itself produced by the preceding $k$ steps, and comes into being  only once those steps have been executed. Rice's theorem \cite{Rice1953} establishes that no general algorithm can decide non-trivial semantic properties of such trajectories regardless of the time available. This does not preclude the existence of some shortcuts (for example, stretches of the trajectory governed by a temporarily stable rule), but in the general case the exact trajectory is essentially temporal. Additionally, where the generation process introduces information not derivable from earlier states (for example, information from the environment), the Kolmogorov complexity \cite{Kolmogorov1968} of the trajectory grows with the number of steps that introduce information, and there is no description substantially shorter than the simulation itself. The trajectory is then incompressible.
}
}.
Hence non-trivial generate and test is essentially temporal, embedded in natural time.

\subsection{Embryogenesis and metamorphosis are iterative processes}
\label{subsec:emb-and-meta}

It makes sense that Turing studied intelligence and computation, but also produced a model of self-organising patterns during embryogenesis. As he saw, the formation of bodies and the formation of minds are closely linked and in many ways isomorphic problems. The cognitive apparatus of human and non-human animals results in an emergent coherent {\it self}: a first-person perspective that has memories, goals, and preferences distinct from the individual neurons making up the brain. 
Long before bioelectric networks became the cognitive glue for the cells in the brain, they served the same purpose in the body.

Each of us began life as a single cell, slowly and gradually completing the journey from the chemistry and physics of an unfertilized oocyte to the second-order metacognition of complex mammals \cite{arango2011two}. How do the individual cells cooperate and compete to complete a specific path in anatomical space, in order to build a functional body? A single embryonic blastoderm might have tens of thousands of cells, and yet we look at it and see `one embryo'. What is there {\it one} of, in this collective? The mechanism that allows the emergence of a discrete individual from this cellular excitable medium is {\it alignment}: the ability of these cells to work as a collective intelligence to achieve a goal state in anatomical morphospace, not merely to exhibit emergent complexity (open-loop pursuit of local rules), but to solve problems (repair injuries, withstand noise and other perturbations of DNA and the environment) and arrive at the correct target morphology despite novel situations.

{In morphogenesis each biological stage builds on the previous one, whether it is embryogenesis, regeneration, or metamorphosis. It is iterative in two key ways. 
First, it involves a kind of de-noising and iterative error minimization relative to a setpoint. 
Regulative development and regeneration progressively expand, delete, and rearrange tissue, \textit{stopping when a specific target morphology is reached} \cite{Levin2023-tele,Levin2026}. It is a homeostatic process that traverses the anatomical space as needed, to achieve a specific state when deviated from that state.} 
Stem cells are born as totipotent cells, and as they become enmeshed in interaction with other cells new generations become pluripotent until they finally are dedicated to one single purpose \cite{shah2021types}. Precursor cells differentiate and migrate so as to fit with their capabilities into the overall organism's needs.
{Second, in the sense that morphogenesis involves progressive edits to modify a system, made by the system itself, it is a remarkable example of self-reference: the very computations needed for the cell collective to know what to do next to minimise distance from their target shape are carried out \textit{by the cell collective itself}. The decision-making medium is re-shaped, as are its goals (for the next developmental stage) and capabilities (because the network architecture grows and changes), by its decisions and actions. It is autopoiesis in the strongest form, and a self-referential control of radical transformation with computation `on the fly'.}

\subsection{Alternative non-iterative processes}
\label{subsec:alternative}

\begin{quote}
    \textbf{Are there non-iterative processes that can effectively explore a Vast state space of possibilities?}
\end{quote}
\noindent
Evolution’s generate and test is essentially temporal, as each generation depends on the previous one.  Growth from a seed to a full organism is also essentially temporal, due to the growth processes at one time depending on and building on the structure present at that time.

Are there physical mechanisms other than generate and test or growth and development, that could be exploited in order to explore the Vast state space of possibilities more effectively, and in a non-iterative manner?  Could, for example, some form of quantum collapse to a viable candidate be possible in a single step?  Or would there still need to be some iterative process to arrange the superposition state into a form where collapse yields a viable candidate with non-negligible probability?  

These are open questions.  But for now, it seems that the evolution and development of life – the discovery of the viable sub-space – are essentially temporal, requiring the passage of natural time.

\section{Self and Self-reference}\label{sec:self}

In this section, we first discuss what is a (living) \textit{self},
and how it can be self-referential.
We then discuss how the concept of natural time can be used to unwind the self-reference,
and so avoid various {apparent} paradoxes.

\subsection{Establishing a Self}\label{sec:establishing}

Living systems are distinguished from other material systems by various remarkable capacities, including self-production, self-maintenance and self-governance (summarised as autonomy by Varela  \cite{varela1979}).   
What is meant by the term `self' in this context?  One route to address this question is to look at how it is that a notional self can come into being, and how that, in turn, leads to self-reference. 

\paragraph{Some-thing from no-thing: the role of the observer.}
The central tenet of second-order cybernetics \cite{mead1968,glanville2013} is that you cannot have observations without observers; this change in emphasis from observed systems to observing systems has led to a (radical-)constructivist epistemology.
As the observer will observe: ``changing, it rests'' (Heraclitus, DK B64). Amidst constant change, there are things in the world that persist, that we recognise, identify and name. Such an object (thing or entity) is a token that refers to a stable pattern (form or behaviour), as discerned by an observer \cite{Von_Foerster2003-dv}. In this way, the observer makes the distinction of some-thing (where previously there was no-thing, or another thing) other than (them)self.

\paragraph{Invariance and recursion, vs iteration.}

To recognise a pattern as an entity is to suggest that it retains a degree of similarity over time: an invariance. That the entity is dynamic, yet invariant, suggests that whatever changes it undergoes, it finds its way back, thereby (re)forming the pattern: its character. von Foerster  termed this characteristic an \textit{eigenform} and the process that gives rise to it \textit{eigenbehaviour} \cite{Von_Foerster2003-dv}.  

Thus the pattern-generating process references  prior cycles in a recursive manner. In this way, invariance and recursion are inseparable and complementary. An alternative view of the same process focuses on the step-by-step process by which the constitutive elements build up the whole pattern through iteration. Thus, recurrence and iteration are complementary views, representing different indications (by an observer) of the part-whole dynamics.

\paragraph{Circularity, causality and closure.}
In recursion we are dealing with processes arranged in a circular manner, i.e. with (topological) closure. Systems with such organisation challenge traditional mechanistic modes of explanation in the form of (linear) causal chains (e.g. A causes B causes C) in which cause temporally precedes effect. In topologically closed recurrent processes, basic notions such as cause and effect, past and future, become problematic, as the process loops back upon itself (or re-enters its own form \cite{spencer-brown1969}); traditional linear causal chains are not present, or appear as artefacts resulting from various cuts (by an observer) of the circular organisation. 
{Figure~\ref{fig:repclosure} sketches four traditions (self-organisation, relational mathematics, control theory, and epistemology) that independently converge on circularity (closed-topology) and self-reference as the structural condition for autonomy, and, in the biological case, for life itself.}

\paragraph{On the notions of self and self-reference.}
If, in observing, the entity appears to be involved in, or constitutive of, its own becoming, the observer may then identify (make the distinction of) the entity as a \textit{self}, and then use the prefix `self-'  to invoke, for example, self-reference. 
The connotation of self-reference is that there is an entity, for which the distinction as a self can be made, that refers to the entity as a whole, and that enters into the processes that the entity (i.e. self) undertakes (and/or of which it is constituted); self-reference can also be thought of as a re-entry.  In this way, self-reference connotes (some degree of) autonomy (for example, self-regulation) of the entity. Varela \cite{varela1979} focused on autonomy (and its origins) as the key distinguishing characteristic of living systems.

\begin{figure}[tp]
\centering
\tikzset{every picture/.style={line width = .3mm, font=\sffamily}
}
\resizebox{0.95\linewidth}{!}{%
\begin{tikzpicture}
\usetikzlibrary {arrows.meta,calc,decorations.markings,decorations.pathmorphing,shapes}

\begin{scope}
    
  \node (org) at (0,1.5) {ORGANISM};
  \draw[
    decoration={markings, mark=at position 0.125 with {\arrow{>}}},
    postaction={decorate}
  ] (0,0) circle [radius=1cm];
  \draw [-{Straight Barb[right]}] (0.15,-2.7)--(0.15,-1.3);
  \draw [-{Straight Barb[right]}] (-0.15,-1.3)--(-0.15,-2.7);
  \draw[decorate,decoration={snake,segment length = 6mm}] (-2,-3) -- (2,-3);
  \node (env) at (0,-3.5) {ENVIRONMENT};
  \node[align=left, font=\rmfamily] (a) at (-3,-3.5) {(a)};
\end{scope}

\begin{scope}[xshift=6cm]

    \node[circle,fill,inner sep=2pt] (fdot) at (-1,0.5) {};
    \node (f) at (-1,1) {$f$};
    \node[circle,fill,inner sep=2pt] (Adot) at (-2.5,-2) {};
    \node (A) at (-2.5,-2.5) {$A$};
    \node[circle,fill,inner sep=2pt] (Bdot) at (0,-2) {};
    \node (B) at (0,-2.5) {$B$};
    \node[circle,fill,inner sep=2pt] (Cdot) at (2.5,-2) {};
    \node (C) at (2.5,-2.5) {$\Phi$};

    \draw [-{Stealth[round]}] ($(fdot)+(-0.3,-0.2)$)--($(Adot)+(0.1,0.2)$);
    \draw [-{Triangle[open]}] ($(fdot)+(0.3,-0.2)$)--($(Cdot)+(-0.1,0.2)$);
    \draw [-{Triangle[open]}] ($(Adot)+(0.2,0)$)--($(Bdot)+(-0.2,0)$);
    \draw [-{Stealth[round]}] ($(Cdot)+(-0.2,0)$)--($(Bdot)+(0.2,0)$);
    \draw [-{Triangle[open]}] ($(Bdot)+(-0.1,0.2)$)--($(fdot)+(-0.1,-0.2)$);
    \draw [-{Stealth[round]}] ($(Bdot)+(0.1,0.2)$)--($(fdot)+(0.1,-0.2)$);
   
    \node[align=left, font=\rmfamily] (b) at (-3,-3.5) {(b)};
\end{scope}

\begin{scope}[xshift=12cm]

    \node[draw,fill=yellow!20,minimum size=0.75cm] (s0) at (-2,1) {$s$};
    \node[draw,fill=blue!10,minimum size=0.75cm] (s1) at (2,1) {$s$};

    \node (zero) at ($(s0)+(0.6,0.6)$) {0};
    \node (one) at ($(s1)+(-0.6,0.6)$) {1};

    \node[draw,fill=yellow!20,circle,minimum size=0.75cm] (c0) at (-2,-0.3) {$c$};
    \node[draw,fill=blue!10,circle,minimum size=0.75cm] (c1) at (2,-0.3) {$c$};
    
    \node[draw,fill=yellow!20,minimum size=0.75cm] (e0) at (-2,-1.5) {$\epsilon$};
    \node[draw,fill=blue!10,minimum size=0.75cm] (e1) at (2,-1.5) {$\epsilon$};
    
    \node[draw,fill=yellow!20,minimum size=0.75cm] (ss0) at (-1,-1.5) {$s^*$};
    \node[draw,fill=blue!10,minimum size=0.75cm] (ss1) at (1,-1.5) {$s^*$};

    \node[draw,cloud,cloud puffs=10,cloud puff arc=120,aspect=2,minimum size=0.75cm, fill=red!10] (x) at (0,-3.2) {$x$};

    \draw [-{Stealth[round]}] (s0)--(c0);
    \draw [-{Stealth[round]}] (s1)--(c1);
    
    \draw [-{Stealth[round]}] (c0)--(e0);
    \draw [-{Stealth[round]}] (c1)--(e1);
    
    \draw [-{Stealth[round]},rounded corners] (ss0) |- (c0);
    \draw [-{Stealth[round]},rounded corners] (ss1) |- (c1);
    
    \draw [-{Stealth[round]},rounded corners] (e0) |- (x);
    \draw [-{Stealth[round]},rounded corners] (e1) |- (x);
    
    \draw [-{Stealth[round]},rounded corners] ($(e0)+(0.2,-0.75/2)$)
        -- ++(0, -0.5)
        -| ($(ss1)+(0,-0.75/2)$);
    \draw [-{Stealth[round]},rounded corners] ($(e1)+(-0.2,-0.75/2)$)
        -- ++(0, -0.6)
        -| ($(ss0)+(0,-0.75/2)$);
    
    \draw [-{Stealth[round]},rounded corners] ($(x)+(-0.1,0.3)$) |- (s0);
    \draw [-{Stealth[round]},rounded corners] ($(x)+(0.1,0.3)$) |- (s1);

    \node[align=left, font=\rmfamily] (c) at (-3,-3.5) {(c)};
\end{scope}

\begin{scope}[xshift=18cm]

\begin{scope}[yshift=0.5cm]
    \draw[
        decoration={markings, 
            mark=at position -0.1 with {\arrow{<}},
            mark=at position -0.4 with {\arrow{<}}
        },
        postaction={decorate}
    ] (0,0.5) ellipse (10mm and 5mm);
    \draw (-1.2,-0.1) rectangle node {Feedback} (1.2, 1.1);
    \draw[fill=white] (-1.2,0.8) rectangle (1.2, 1.1);
    \draw[
        decoration={markings, 
            mark=at position 0.15 with {\arrow{>}},
            mark=at position -0.1 with {\arrow{>}},
            mark=at position -0.3 with {\arrow{>}}
        },
        postaction={decorate}
    ]  (-2,0.8) -- (2,0.8);
    \node (in) at (-1.2,1.3) {Input};
    \node (out) at (1.2,1.3) {Output};
    \node (first) at (0,-0.4) {\textbf{Engineer}};
\end{scope}

\begin{scope}[yshift=-1.8cm] 

    \draw[
        decoration={markings, 
            mark=at position -0.1 with {\arrow{<}},
            mark=at position -0.4 with {\arrow{<}}
        },
        postaction={decorate}
    ] (0,0.2) ellipse (20mm and 12mm);
    \draw (-2.2,-1.6) rectangle (2.2, 1.4);
    \node (fb2) at (0,-0.6) {Feedback};
    \draw[fill=white] (-2.2,0.8) rectangle (2.2, 1.4);
    \node[] (second) at (0,-1.3) {\textbf{Weiner, Bateson, Mead}};
    
    \draw[
        decoration={markings, 
            mark=at position -0.1 with {\arrow{<}},
            mark=at position -0.4 with {\arrow{<}}
        },
        postaction={decorate}
    ] (0,0.5) ellipse (10mm and 5mm);
    \draw (-1.2,-0.1) rectangle node {Feedback} (1.2, 1.1);
    \draw[fill=white] (-1.2,0.8) rectangle (1.2, 1.1);
    \draw[
        decoration={markings, 
            mark=at position 0.2 with {\arrow{>}},
            mark=at position -0.2 with {\arrow{>}},
            mark=at position -0.35 with {\arrow{>}}
        },
        postaction={decorate}
    ]  (-2.5,0.8) -- (2.5,0.8);
\end{scope}

    \node[align=left, font=\rmfamily] (d) at (-3,-3.5) {(d)};
\end{scope}

\end{tikzpicture}
}
\caption{
Representations of {organisational closure across four formal traditions.} 
(a) Autopoietic system of Maturana and Varela (after \cite[p74]{Maturana1987}): 
{the organism maintains a topologically closed network of production, continuously regenerating its own components and boundary, co-determining itself and environment.}
(b) $(M,R)$-system of Rosen (after \cite[fig.10C.6]{rosen1991life}):
{metabolism $M$ and repair $R$ processes stand in a loop closed to efficient causation, making the system an irreducible cause of its own organisation.} 
(c) Coupled regulators (redrawn from \cite[p.25]{McNeil2004}): 
{two interacting regulatory sub-systems mutually constrain one another, constituting circular organisation through their coupling.} (d) First-order cybernetics (top) contrasted with second-order cybernetics (bottom), proposed by Mead, Bateson (after \cite{Brand1976}): 
{the former positions the observer outside the system; the latter folds the observer's own act of distinction-making back into the account.} 
}
\label{fig:repclosure}
\end{figure}

\paragraph{Autopoiesis: self-generation.}

Autopoiesis \cite{maturana1991} originates in a conceptual model of the cell as viewed from the level of its molecular constituents. Autopoiesis makes a distinction between organisation (abstract pattern of relationships) and structure (the physical embodiment of the pattern of relationships) and concerns itself with organisation. Autopoiesis identifies the cell as a self-generating (self-maintaining) entity that possesses ``organisational closure''. Organisational closure relates to the circular topology of a network of production (and transformation) processes of the physical constituents of the cell. 
For an autopoietic system, ``the product of its operation is its own organization'' \cite{maturana1991}.

Central to autopoiesis is the notion of {\it structural coupling}: the dynamic process of (mutual) structural changes as a result of interactions between a system and its environment, for example, creating and maintaining a niche. This process leads to the dynamic mutual determination of self (i.e. the cell) and non-self. Here the non-self is not some fixed, pre-existing environmental milieu, but is a world that is brought forth by the self through the enactive dynamics of structural coupling.

The process of structural coupling can be viewed as a trajectory where the living entity makes structural changes as a result of its interactions with the environment (and vice versa) in order to maintain itself in what is a process of mutual adaptation (or learning). These structural changes, in turn, influence its interactions. In this way the entity is \textit{structurally determined}: its identity and behaviour are a result of its history of structural coupling.

The structural determination of living systems, resulting from organisational closure, leads to behaviour which is autonomous, self-determined, in that it ``subordinates all changes to the maintenance of its identity'' \cite{varela1979}.

\subsection{Self-reference and natural time}
\label{subsec:existence}

\begin{quote}
    \textbf{Does the possibility of self-reference in a system require or imply the existence or use of natural time?}
\end{quote}

\noindent
{What has self-reference to do with time, and with natural time in particular? To understand this relation, we need to come back to the issue of the potential contradiction that self-referentiality might cause. Earlier we spoke of paradoxes in the context of conventional logic, which is characterised by an absence of a notion of time. Here we argue that this apparent absence is a result of a conceptual simplification
in conventional logic that is not the case in natural systems.  }

\begin{figure}[tp]
\centering

(a)~\includegraphics[trim = 2.7cm 0cm 2cm 2.5cm, clip, scale=0.8]
{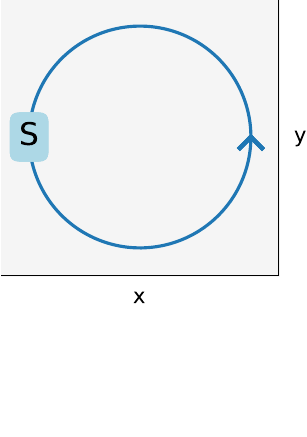}
\hspace{1cm}
(b)~\includegraphics[trim = 0.6cm 0cm 1.2cm 0cm, clip, scale=0.8]{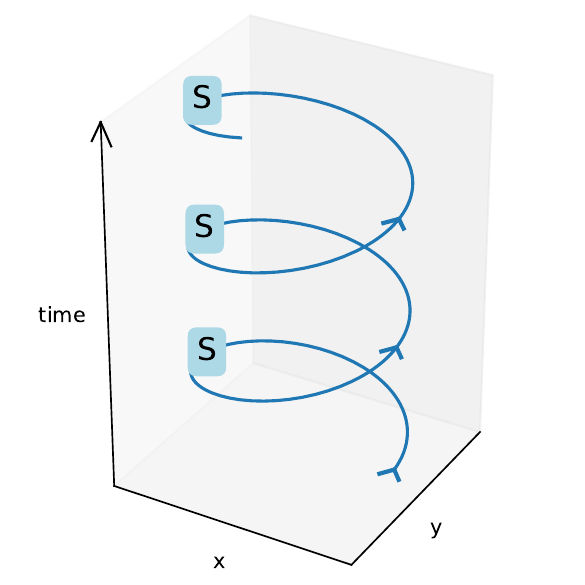}
\caption{(a) `Top-down' view on a system S that refers to itself, symbolised by the arrow; 
(b) adding the dimension of time allows self-reference to be seen as a reference to another system, though possibly similar in every property except the time at which it exists.
}
\label{fig:SR-view}
\end{figure}

Let us first focus on natural systems. If we consider a natural system that refers to itself, how can apparent paradoxes of self-reference be avoided?
Figure~\ref{fig:SR-view}(a)  depicts a system that refers to itself, a special (limit) case of circular reference of one system referring to another, which in turn refers to another until the last system refers to the first. 
In any case, the system referred to last is supposed to be identical to the first. 
But how can that be? The very act of reference (or self-reference) will change the system, even if imperceptibly, since otherwise it could not be said to reference to anything. 
In other words, a system referencing something is distinct from a system not referencing something, otherwise the act of referencing would
be naught.

{In the figure, we symbolise the act of referencing by an arrow, since it is an operation that the system executes.} Operations are by definition activities that have to happen in (natural) time. 
Thus the situation is not correctly depicted in Figure~\ref{fig:SR-view}(a) -- or rather, it is depicted as a projection -- and is instead more accurately represented by the situation depicted in Figure~\ref{fig:SR-view}(b), with the additional dimension of time made visible.
Hence a reference made by a \textit{present} system to itself is resolved by a {\it future} self, 
which unwinds the circle of the apparent paradox into a natural time spiral, hence removing a potential paradox. In fact, we might state that the paradox is an illusion, brought about by a special projection of the situation, the flattening to a timeless situation, which ignores the axis of time.

If the self is changing over time, what does \textit{self}-reference really mean? We assume here that the self is the system under consideration. What we have to conclude from the previous discussion is that the self, while in most aspects perceived the same, is never the same at different times. Even if it has not visibly changed, it can be time-stamped with natural time at any given moment, and will therefore be different by this very property.
{Keep in mind that we are still talking about natural systems.}

{We now move into the conceptual realm.} 
We are accustomed to think that the proof of a mathematical statement is timeless. 
{That is the fascination and power of a proof: It can show that a certain statement is (was, will be) true for ever. }
But indeed, while the proof {\it looks} timeless, it is not: A proof is an operational procedure that starts from assumptions, applies inferences stepwise and ends in the production of a result. 
The \textit{result} is a timeless formal object, 
{TRUE or FALSE, captured in}
a tree of inferences, whose time-dependence can be projected away, but which would not come about without the time-dependent operations of creating it.
{If we look only at the result, we can perceive it as timeless, but what we are really doing is projecting it into an abstract space devoid of a notion of time.}

Are there cases where this operation of projection cannot be executed? 
That is, are there cases where one cannot separate the process aspect of generating the result from the result itself? Following G\"odel’s incompleteness theorem, is this what Mathematics is? More generally, is Science in a similar position?

An example of this temporal unwinding of a timeless projection can be found in the
work of philosopher Patrick Grim and colleagues
\cite{mar1991pattern,grim1993,grim1993self,grim1997}. Grim started with the basic Liar Paradox (``This sentence is false'') and pointed out that it is a paradox only if one insists on a single constant (i.e., timeless) truth value. 
If one places this self-referential sentence in what we call natural time,
unwinding the circle into a helix as in figure~\ref{fig:SR-view}, then it can be seen to encode a simple oscillator that bounces between TRUE and FALSE through time. By using fuzzy logic to assign truth values on the interval from 0--1, he then showed how systems of mutually self-referential equations (e.g., ``Sentence X is twice as true as sentence Y is'' etc.) can be plotted as dynamical systems with interesting behaviours and morphologies.

{Analogous considerations can be used to resolve other paradoxes, e.g. the question of \textit{causal conflict} in systems where emergence plays a role \cite{kim2006}: By introducing a notion of time, they suddenly disappear.}

\section{Time, Life, and Self-Reference}
\label{sec:TLR}

Temporality and the self-referential nature of life are most clearly seen in the processes of \textit{morphogenesis}: the generation of complex form during embryonic development, regeneration, or metamorphosis. The process is not merely complex (emergent, open-loop) but actively robust, being able to traverse to the same outcome in anatomical morphospace despite different starting conditions and various perturbative changes in environment and even cellular composition. This requires a system to perform anatomical homeostasis: actively taking measurements, storing setpoints, and working to reduce the error. Such homeostatic loops inherently define arrows of time for the organism where stress rises and falls, in part due to the decisions it makes.

The remarkable thing about self-assembling biology is that there is no material separation between the mind and the body: the cell networks making decisions about what to do next in their journey through morphospace are the very same tissues that are being actively remodelled (see also footnote~\ref{foot:no-separateion}). The computational medium and its work product are the same, thus implementing a remarkable ``strange loop'' \cite{hofstadter2008} in which the computational competencies and the meaning of all the signals and state information is constantly changing because the machine itself is being actively remodelled in real time.

Morphogenesis is self-referential in the deepest sense, because this proto-cognitive system is constantly {\it thinking about itself} and how to change itself to achieve specific goals and reduce stressors that indicate delta from the next stage.  
 
Even single cells have the above two features, but they deploy them at smaller scales and in other spaces besides anatomical morphospace, by navigating in physiological, transcriptional, and metabolic problem spaces. In all these cases, active inference and other processes require the system to have a model not only of the outside world, but of itself;
the fundamental loop driving these changes involves prediction/perception/action cycles that fundamentally involve self-reference and are grounded in a notion of time and the drive act to reduce surprise and error.

\subsection{Embryogenesis: developing the boundary between self and world}
\label{subsec:embryogenesis}

Anatomical problem-solving is a type of intelligence, in William James’ definition \cite{James-1890}. The emergence of agents capable of pursuing goals in behavioural, anatomical, and physiological spaces requires an integration of information processing and a scale-up of the set-point that drives homeostasis. This integration also requires solving the problem of the boundary: every cell has other cells as its neighbours; where is the boundary between self and world? 

We can watch embryos solve this existential issue by making scratches in the blastoderm: each island will become its own embryo, leading to a set of conjoined twins, triplets, and so on \cite{McMillen2024}. This means that the number of coherent beings, or \textit{selves}, in an embryonic blastoderm is not fixed by genetics but can be anywhere from zero to probably half a dozen, arising in the physiological process that demarcates subsystems that all cooperate toward a specific goal.
An embryo is fundamentally a self-reinforcing story, a model that becomes increasingly more convincing to cells as they implement the journey in anatomical space to which it binds them. It is fundamentally a self-referential process of circular causation in which the larger scale of organization distorts the option space for its competent parts, and their actions in that space build and reinforce the existence and coherence of the higher level form.

The fact that living agents are built on a multi-scale competency architecture has other implications. Individual cells come and go all the time, as part of normal homeostatic turnover, but the large-scale structure of the body resists ageing and decay for decades. This \textit{Ship of Theseus}
(an object that remains the `same' object despite having all its components replaced)
is defined by the bioelectric, biomechanical, and biochemical pattern memories in tissues that guide the repair machinery and maintain a species-specific target morphology over time despite considerable noise at the lower levels. This process is self-referential in an important way, because the cells keeping the memory of what to build are the same cells which are implementing that pattern by moving and replacing themselves and other cells, all at the same time.

Our bodies feature top-down control in which large-scale executive behavioural goals of our cognitive selves filter down to control ion flows across muscle cell membranes, for example in the case of voluntary motion undertaken to meet highly abstract social goals. This is but one perspective of a myriad of regulatory feedback systems that span multiple levels of organisation, an idea formulated as ``no privileged level of causation'' \cite{noble2012}.

The same multiscale control system is used in the morphogenetic collective intelligence, where coarse, large-scale bioelectric set-points are transduced into changes of individual cell behaviour and gene expression, to implement a desired path in anatomical morphospace
(this concept evokes Waddington's concept of developmental canalisation \cite{Waddington1942}). 
This architecture is what makes living things (but not yet our current single-level engineered devices) susceptible to cancer: defections of parts who disconnect from the collective, revert to their ancient evolutionary past as microbes with tiny cognitive \textit{light cones}
(a functional boundary that defines the scale and limits of a self's cognition
 \cite{Levin2019})
and tiny goals, and treat the rest of the body as just external environment (a shrinking of the border between Self and world).

\subsection{Metamorphosis: maintaining the self through radical change}
\label{subsec:metamorphosis}

The flow of information within this architecture is fascinating and enables behavioural robustness that we have yet to even begin to emulate in silico \cite{levin2023,levin2023a}. 
The caterpillar can be trained, and the butterfly or moth will retain those memories, despite the fact that the brain  is almost totally destroyed and rebuilt during the metamorphosis process \cite{Blackiston2008,Blackiston2015}, and that those memories have to be decoded in the context of a totally different body structure and behavioural repertoire, from a soft-bodied creature that lives in a 2-dimensional world and eats leaves, to a hard-bodied architecture that flies through a 3-dimensional world and drinks nectar. Similarly, planarian flatworms regenerate their entire brain and their memories after their heads are amputated \cite{Corning1967,Shomrat2013,Blackiston2015}.

While these examples seem like unique features of {so-called} lower life forms, in fact they highlight a universal dynamic. We humans  do not have direct access to the past: at any moment, we must actively reconstruct our past history (models of self and world) from physical engrams left for us by our prior experiences, 
which constrain and modify the energy landscape of possible actions for our future selves. 
We are thus a stack of \textit{selflets}, in a continuous dynamical process that maintains and remodels both our bodies and our minds by taking advantage of the plasticity of our multiscale nature as collective intelligences operating in diverse problem spaces \cite{varela1991}. 
Crucially in biology, memories (whether behavioural engrams or genomic sequences) must be interpreted: a dynamic that eventually scales cell-level problem-solving into the advanced intelligence we observe in brainy animals \cite{Levin2024}. 
The precise dynamics necessary and sufficient for 
creative interpretation of physical memory structures in a way that facilitates forward-looking dynamic goals remains
very poorly understood, but it is clear that electrophysiology of all cells, not just of nerves and muscle, is centrally involved.

Selves are fundamentally processes in constant flux. At every level, material and cells are being replaced, yet the whole persists. The Ship of Theseus that is our body remains, because persistent information structures guide cell behaviour toward a kind of large-scale anatomical homeostasis. 
An agent is just a model in the mind of some other agent; a Self is a model of an agent in that agent’s own mind. 
{Living things}
 model themselves as agents 
{because they}  cannot afford to track microstates.
Survival requires being good at coarse-graining reality into manageable chunks, and telling agential stories about those chunks in the world, which are then tractable for real-time decision-making. Thus, we were committed to an agency-first view of the world 
from the microbial stage, because of the impracticality of a micro-reductionist worldview for survival in the real world.

Once a system is good at formulating agential models of events in the outside world, it can turn that perspective onto itself, in effect become self-aware, realizing that it too is an agent that does things. This closes the loop on the fundamentally self-referential process, in which evolutionary pressures to survive despite metabolic scarcity and time constraints drives the appearance of models of the Self as a persistent, unified being.

\subsection{Learning and modelling the self}
\label{subsec:learning}

Dennett \cite{dennett1996} introduces a {\it tower} of generate and test algorithms.  At each level of the tower, the generation process becomes more sophisticated, from Darwinian mutation, through Skinnerian reinforcement learning, Popperian predictive models, to Gregorian social learning through culture.  Ever more intelligent or creative generative methods, along with ever more intelligent selection methods, can help move through this Vast search space more effectively.  In higher levels of the tower, these methods involve more explicit cognitive memory, and self-reference: learning, building and using ever-improving predictive models of the environment and self \cite{Clark-surfing-2016}.  As creative humans, we can further apply our intelligence to develop technologies to accelerate the processes of natural evolution (artificial selection, genetic engineering, xenobiology, etc.).  In this way, the underlying processes of evolution (the generate and the test stages) can themselves evolve, leading to meta-evolution, and potentially more effective ways of exploring the Vast state space.

\subsection{Self-modification}

We have noted that several paradoxes can be resolved by turning the timeless circle of self-reference into a spiral in natural time
(section~\ref{subsec:existence}).
In reality, nothing is a perfect circle:
other changes during the cycle result in growth, difference, amendment.
One important aspect of this is self-amendment:
when the self changes itself, whilst remaining itself.

Suber \cite{Suber-1990} addresses this topic from the context of (social) law.
The legal system includes legal rules that govern the change of other legal rules,
and even has higher level rules for changing the rules for changing the rules.
Suber \cite[preface]{Suber-1990} points out that ``The paradox of self-amendment arises when a rule is used as the authority for its own amendment.
It is sharper when the rule of change is supreme, sharper still when it is changed into a form that is inconsistent with its original form, and sharpest of  all when the change purports to be irrevocable.''
Of course, these are social rules, not natural laws, and so can be broken, but potentially at huge cost.

As part of his exercise in investigating self-amending rules,
Suber \cite[App.3]{Suber-1990} introduces the game\footnote{%
Suits \cite{Suits-2014} defines a game as ``the voluntary attempt to overcome unnecessary obstacles''.
Nomic is a game, as playing it is voluntary;
the legal system is not a game, as `playing' it is enforced.
}
of Nomic,
``a game in which changing the rules is a move''.
Its first rule, Rule 101, is to obey the rules.  
Although this may seem redundant, Suber points out that ``Rule 101 is included precisely so that it can be amended; if players amend or repeal it, they deserve what they get.''
Rule 208 states that ``The winner is the first player to achieve 100 (positive) points''; of course, this rule may also be changed.
The initial setup for Nomic is thus of a \textit{finite game}, where there is an end point and a winner. Carse \cite{Carse-1986} notes that ``A finite game is played for the purpose of winning, an infinite game for the purpose of continuing the play'' -- the contrast between a final \textit{destination} and an ongoing \textit{journey} --
and that infinite games are more open, playful and creative.

Suber notes that a game of Nomic can get very complex as it proceeds,
but that attempting to develop a computer program to keep track of the state of the rules, and to determine the validity of a move,
is as difficult as playing the game itself:
the program would need continual modifications to keep up with the changes.
It is possible to write programs that change rules, at least at one level.
The computer game \textit{Baba is You}\footnote{%
\url{https://store.steampowered.com/app/736260/Baba_Is_You/}
} is ``a puzzle game about changing the rules'',
somewhat less complex than Nomic, at least initially.

\subsection{Open endedness}\label{sec:open}

Self-modification is a route to open endedness, the continual generation of  novelty, through evolution, metamorphosis, and other mechanisms.
Open endedness is necessary in any system that does not merely recur: it opens the timeless cycle not merely into a temporal helix (fig.\ref{fig:SR-view}), but into a growing spiral.

In order to capture the concept of open endedness,
it is necessary to have an appropriate definition of `novelty'.
Banzhaf et al. \cite{Banzhaf-2016} provide such a definition of  novelty, with respect to models and metamodels of the system (see figure~\ref{fig:metamodel}).
They identify three levels involved in systems:

\begin{figure}[tp]
\centering
\includegraphics[scale=0.6]{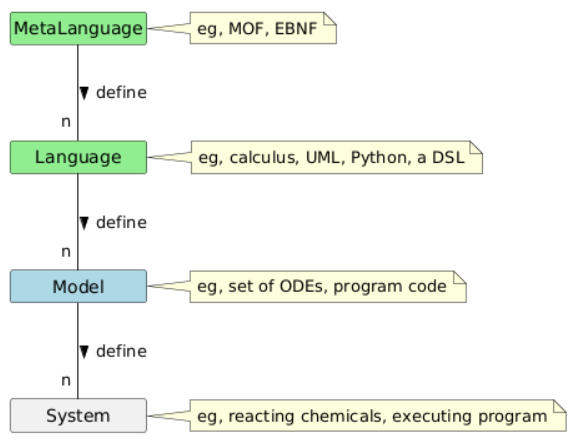}
\caption{The relationship between systems (physical things in the world, or simulations), the model that describes or defines them, the language (or metamodel) used to define models, and the metalanguage used to define languages.
}
\label{fig:metamodel}
\end{figure}

\paragraph{Level 0: system instances.}
These are the physical instances in a real world system, such as a collection of reacting chemicals,
or the virtual instances in an executing computational system, such as a collection of interacting virtual agents.
Instances can change over time:
they may move, they may change state.
These changes are termed \textit{variation}, or level-0 novelty.

\paragraph{Level 1: system model.}
The system model describes or defines (depending on whether it is a descriptive or prescriptive model) the allowed instances and their behaviours over time: it captures the relevant state space and dynamics.
It may be a descriptive scientific model, for example describing how a set of chemicals react;
it might be a prescriptive engineering model,
for example defining how a collection of virtual agents interact.
Models themselves are typically static (have no dynamics): they exist unchanged while their instances have dynamics.
However, if the model does change for any reason,
does exhibit dynamics, then this kind of change is termed \textit{innovation}, or level-1 novelty,

\paragraph{Level 2: metamodel, or language.}
Models themselves are things in the world, and so can be modelled.
The model of a model is called a metamodel;
it captures the concepts used in the model.
The metamodel can define a language for expressing models;
if so, it is typically identified with that language \cite[sec.8.1]{Kleppe-2003},
and the model is written in the language.
For example, the model of a set of reacting chemicals might be written in the language of calculus as a particular set of differential equations;
the model of interacting agents might be written in a programming language, as a specific program code in that language.
Metamodels themselves are also typically static.
However, if the metamodel does change for any reason, then this kind of change is termed \textit{transformation}\footnote{%
originally called \textit{emergence} in \cite{Banzhaf-2016}, but renamed \textit{transformation} in \cite{Stepney:2021:oee4}.
}, or level-2 novelty.

\vspace{2mm}
\noindent
Banzhaf et al. \cite{Banzhaf-2016} consider variation to be a normal property of any dynamical system, and so not the kind of novelty of interest in open ended systems.
So they define an open ended system to be one that continually produces type-1 and type-2 novelties.
There is conceptually the possibility of type-3 and higher level novelties \cite{Stepney:2021:oee4}.

Models are  static {when} their metamodels have no concepts, or language features, available for changing the model, they simply provide the syntax for writing models, and the semantics of the resulting model.
For example, calculus provides no mechanism for adding new equations to a set of ODEs as they are being solved, or even for recognising the need for such a change.
Similarly, (most) programming languages provide no mechanisms for changing code as it executes:
indeed, they are designed to prevent this very possibility.
Similarly, metamodels are  static {when} their metalanguages have no features to allow their dynamic modification.
This suggests {a} route to modelling open ended self-modifying systems, including living systems, is to use  models and languages that themselves have self-modifying capabilities (see \cite{Banzhaf-2016,Stepney:2023} and section~\ref{sec:comp-tools:self-mod}).

\section{Tools for modelling self-referential systems}
\label{sec:tools}

\begin{quote}
    \textbf{What are appropriate mathematical formalisms and computational tools for modelling a self-modifying system that changes its own rules, and is open to its environment?}
\end{quote}
\noindent
As we have argued above, consideration of self-reference is crucial to understanding living systems, and the consideration of time is crucial to resolving some of the apparent paradoxes of self-reference.
In this section we discuss some of the requirements for mathematical formalisms and computational tools for modelling, simulating, and analysing such systems.
{We also give a set of examples of tools and models, both those developed by some of the coauthors, and those from the wider literature, that we deem appropriate for use in modelling and analysing aspects of living systems.
Application of these approaches to realistic living systems is still an open research problem.}

\subsection{Mathematics of self-reference and self-modification}\label{sec:maths}

The ability of symbolic systems to refer to themselves is not merely a marginal feature that gives rise to annoying  paradoxes.
Rather, it is a central feature in the development of modern logic and the mathematical theory of computation. Moreover, this feature arises in the foundations of biology and cognition, and also in game theory and economics.
We give a brief, non-technical overview of some of these developments.

\subsubsection{Taming self-reference}\label{sec:taming}

A seminal moment in the program of developing rigorous foundations for mathematics came with Russell’s paradox, which showed (in modern terminology) the \textit{inconsistency of naive set theory}: {that the informally defined set theories used to that point, which allow arbitrary properties to define sets, could lead to logical paradoxes.}
This, and a number of related paradoxes, led to the development of various schemes that have dominated subsequent discussions of foundations of mathematics. The motivation for these schemes is essentially to \textit{avoid} the paradoxes, whether by limiting the principles for forming sets, the logic used in mathematical reasoning, or both.

Russell’s paradox and its many avatars all involve self-reference in one form or another, whether overt or implicit.
Famously, Russell’s paradox involves the set $R = \{ x \mid \neg (x \in x) \}$, the set of all things that are not elements of themselves. Note the dual role of $x$, as both a set and an element of that set, and the familiar cyclic reasoning this gives rise to when asking ``is $R$ an element of itself?''.
The basic strategy of Zermelo-Fraenkel set theory, the `industry standard’ axiomatic scheme, is to avoid such situations by allowing sets to be built up only in a well-founded, tree-like fashion.

However, as subsequent developments have shown, the issue of self-reference is not so easily avoided.
G{\"o}del showed that as soon as axiomatic theories acquire a modicum of expressive power -- a modest amount of arithmetic is all that is needed -- then one can encode the machinery of self-reference sufficiently well  to be able to represent a syntactic version of the Liar paradox \cite{Goedel1931,Meltzer1962}. The only escape clause from actual inconsistency of the system is \textit{incompleteness}: the inability of the system to prove certain assertions either true or false.

We also find self-reference in the closely related foundations of computation being developed around the same time as G{\"o}del's incompleteness theorem. The \textit{undecidability of the Halting Problem} 
{(that no algorithm can exist to determine if any arbitrary computer program halts or executes forever \cite{Turing1937})}
hinges on a machine being applied to a description of itself, a clear analogue of the Russell paradox, and the self-referential G{\"o}delian sentences.
A functional form of this can be found in Church's $\lambda$-calculus, {which has} the ability to form expressions such as $xx$, where $x$ (as function) is applied to itself (as argument). This is the functional analogue of the self-membership $x \in x$ in Russell’s paradox.
The $\lambda$-calculus, in this unrestricted form, is perfectly consistent, and indeed forms the basis of programming languages such as {SCHEME.}
It also gives a characterisation of computability equivalent to the Turing machine model.

The essential price to be paid, as with any universal model of computation, is that we must accept that some of the functions we describe will be \emph{partial}: the corresponding computations will get stuck in a loop, and go on for ever without returning a result.
If we wish to curb this possibility of infinite loops, we must follow a strategy analogous to that used in set theory: we must restrain the application of functions, by introducing types. This leads to modern typed functional languages such as Haskell, and to type theories such as Homotopy Type Theory \cite{Aczel-HTT-2013}.

\subsubsection{Embracing self-reference}\label{sec:embracing}

In this article, however, rather than wishing to tame self-reference in the name of foundational security and consistency, we wish to embrace it, pursuing the idea that life and cognition involve self-reference and self-modification in an essential fashion.

The mathematics of self-reference has been developed within the theory of computation in the areas of \emph{domain theory} and \emph{coalgebra} (among others) \cite{abramsky1994domain}. We  briefly discuss each of these, and then turn to applications of these ideas in the foundations of biology and of economics.

\paragraph{Domain theory.}

The founding construction of domain theory by Dana Scott was precisely to give mathematical sense to the self-application $xx$ which 
was at the root of the expressiveness of the untyped $\lambda$-calculus. To do this, he constructed a {mathematical} space $X$ which was isomorphic to its function space $X \rightarrow X$. Thus, via this isomorphism, each element of $X$ could be seen simultaneously as an argument, and as a function applicable to any element of $X$, and in particular to itself \cite{scott1970outline}.
No such spaces had been seen or expected in mathematics previously.
This led to a rich development of a theory of domains closed under many mathematical constructions, as a foundation for the semantics of programming languages \cite{gierz2003continuous,abramsky1994domain}.

There is a certain price to pay: 
{the same need to accept partial functions as described in section~\ref{sec:taming}.} 
This may limit their potential applicability in various modelling situations, as we shall see.

\paragraph{Coalgebra.}

Another important development has been the rise of \emph{coalgebra},
a strikingly novel development.
Mathematical induction and algebra have been staples of mathematics for centuries.
There is a debate in the historical literature as to whether induction was known in ancient Greek mathematics.
Algebraic structure is expressed in terms of \emph{operations} that build new values from old values. The most familiar examples are addition and multiplication in arithmetic, abstracted to the operations in groups and rings. Induction goes alongside algebra, as a method of proof for {so-called} \emph{free} algebraic structures, where the elements are exactly those built by application of the operations, starting from some basic constants. For example, the usual mathematical induction on the natural numbers concerns properties that are true for zero, and preserved by the \textit{successor} {(add one)} operation. Since all natural numbers can be built by applying the successor function repeatedly to zero, such properties are true for all of them.
Induction and algebra in this sense represent a bounded universe, where everything is built from below, {from the basic constants}, in a well-founded fashion. This is closely related to the way that the standard universe of axiomatic set theory is built from below.

Much more recent arrivals on the scene are the duals to these notions: \emph{coalgebra} \cite {rutten2000universal,jacobs2017introduction} and \emph{coinduction} \cite{kozen2017practical}. Rather than operations that build or \textit{compose} new elements in a closed form using previously given, closed form elements, we have `co-operations’, which \textit{decompose} elements, in a potentially unbounded fashion.
This in turn requires a novel principle of proof, namely coinduction, which provides an elegant means for reasoning about the processes generated by unfolding coalgebras.

One can, roughly speaking, see coalgebra as sitting between standard set theory and domain theory.
It sits close enough to set theory that it can be seen as an extension of it: indeed, a particular form of coalgebra corresponds to non-well-founded set theory \cite{barwise1996vicious}, which admits objects such as $x = \{ x \}$, but not the Russell set. This means that it is well-adapted to a wide range of modelling tasks, where standard set-theoretic descriptions are extended to accommodate potentially unbounded infinite behaviours.
On the other hand, it cannot capture the full extent of self-reference, as manifested e.g.~by the untyped $\lambda$-calculus.

Simple examples of objects in a final coalgebra are (infinite) \textit{streams}.
The stream comprising the natural numbers $\langle 1,2,3, \ldots, n, n+1, \ldots \rangle$ has no recurrence.
The self-referential equation $S = \langle a, b, S \rangle$ generates the recurrent stream $\langle a,b,a,b,a,b, \ldots \rangle$. We can think of the equation as defining a cyclic system: ``do $a$, do $b$, then we are back where we started''; the stream is the result of unwinding the self-referential equation in time.

\paragraph{Calculus of Indications.} 
Another approach to self-reference emerges from Spencer-Brown's Laws of Form \cite{spencer-brown1969}, which develops a Calculus of Indications 
{that begins with a single primitive act of closure, at once the drawing of \emph{distinction} and the act of \emph{indication}, of marking one side, constituting the other as unmarked, producing the \emph{form}. Writing $\markSB{}$ (pronounced {`cross'}) for the act of marking, the calculus is governed by two axioms. 
The axiom of \emph{calling}, $\markSB{}\;\markSB{} = \markSB{}$, says that repeated indication of the same state yields that state. 
The axiom of \emph{crossing}, $\markSB{\markSB{}} = \quad$, says that crossing a distinction twice returns to the original (unmarked) state. 
These axioms generate the primary arithmetic upon which a complete formal algebra can be built, in which variables may take either the marked or unmarked value. This algebra serves as a precursor to Boolean algebra, although with the crucial difference that $\markSB{}$ functions simultaneously as both operator and operand.}

{The system's significance for self-reference emerges in expressions of the type $J = \markSB{J}$\,, where the form repeatedly folds back upon itself (crosses its own boundary) and cannot settle on a fixed point. To see why: if $J$ is taken as marked, the crossing yields unmarked; if taken as unmarked, the crossing yields marked. Neither assignment is consistent; the expression is irresolvable within the two-valued calculus, generating instead a pattern of alternating marked and unmarked states. 
Spencer-Brown termed this indeterminacy an \emph{imaginary value}, a third state in addition to the marked and unmarked; his insight was to see this imaginary value as real in relation to time. In extension, Varela \cite{varela1975calculus} developed a three-valued calculus with the imaginary value formalised as an \emph{autonomous} state $\markVK{}$ (pronounced {`self-cross'}), arising from self-indication (or re-entry into the form). The interpretation of $\markVK{}$ is complementary: as both a description in space (a vibration), and a prescription in time (a temporal oscillation), with $\markVK{}$ acting as the operator for the recursive dynamic $J \rightarrow \markSB{J}$\,. Thus, the temporal unfolding of the recursive dynamics of closed self-referential systems is internally constitutive, rather than externally parameterised.}

{The view extends from the formal to the biological. Varela's extended calculus shows what it means for recursivity and distinction to be mutually constitutive: the autonomous value is not imposed but emerges. Its constitutively temporal character is
developed in detail with Kauffman in their \emph{Form Dynamics}
\cite{kauffman1980}. For Varela, this mutual constitutivity is the abstract signature of autonomy, of which autopoiesis is the molecular instantiation: the cellular network whose recursive processes bring forth the very organisation they constitute (section~\ref{sec:establishing}).
The Closure Thesis states the generalisation directly: \emph{every autonomous system is organisationally closed}, with eigenbehaviour constituting the invariant identity (eigenform) of any such closure \cite{varela1979}. Kauffman's subsequent
mathematical development of eigenform has consolidated and deepened this shared programme across formal, biological, and cognitive domains \cite{kauffman2005}, revealing \textit{recursivity constituting distinction} as a universal condition of identity-sustaining systems, and, as such, the formal ground of self-reference itself.}

\subsubsection{The mathematical future}

With the further developments in applied category theory \cite{fong2019invitation}, type theory \cite{pelayo2014homotopy}, probabilistic and differential programming \cite{staton2016semantics,heunen2017convenient,huot2023omegapap},  etc. which we have seen over the past decade, the time may now be ripe for  renewed attempts to advance these programs of developing rigorous mathematical foundations for reflexive biology and behavioural sciences.

\subsection{Computational tools}\label{sec:comp-tools}

\begin{quote}
    \textbf{What are effective tools and modelling paradigms for implementing and running (simulations of) self-modifying open systems?}
\end{quote}
\noindent
{Some tools and approaches for modelling self-referential and self-modifying systems are summarised here,
then we discuss tools to analyse self-modifying systems, and end with a discussion of applications. These tools address self-reference and self-modification from distinct intellectual traditions, united by a common aim: to model systems whose organisational rules, and not merely their states, change over time.}

It is no coincidence that {some of} the mathematical formalisms above had their genesis in computational science.
Self-referential systems are natural in computing;
self-modifying ones less so, but not unknown.
{Time is a required component of all computing, as computer programs have not only to be written, they have to be interpreted by a machine and executed. This can only be achieved in natural time, as operation after operation requires time. There is no shortcut, just like phenomena in living systems, which also require natural time for them to happen. That is why}
computer science gives us {more appropriate} tools for modelling, simulating, and analysing such systems,
{in particular their self-referential or self-modifying aspects.
Mathematical approaches such as MES employ categorical frameworks, addressing hierarchical emergence and genuine temporal dynamics. Each approach advances the modelling frontier, while the deeper challenge identified throughout this work, of modelling systems whose self-referential dynamics generate genuinely new domains rather than merely accommodating them, remains open.}

\subsubsection{Genetic Programming: Evolving models}\label{sec:gp}

Genetic Programming (GP) \cite{koza1992} is a technique gleaned from natural evolutionary processes as found in the living world \cite{darwin1859}. It works on a population of computer programs and generates computer code with stochastic mechanisms like code mutation and code recombination. These variations are subjected to a selection process that judges the programs based on quality criteria called `fitness' that are supposed to capture the ability of those programs to fulfil the function envisioned by the outside user. The process is implemented in an iterative loop that tries to improve the fitness of programs until a termination condition is reached. The entire process is executed in a safe computational enclosure such that the modification of code cannot reach code outside the enclosure. 
What is being evolved is a program, that is, a \textit{model} of some behaviour as captured by data (fig~\ref{fig:metamodel}), rather than a mere \textit{instance} (a data point),  although the process {of mutation and recombination} happens outside the model itself. 

The power of evolution becomes more visible, however, if the code inside the safe enclosure is allowed to self-modify. This can happen, e.g. if the stochastic operators that manipulate individuals of the code population are themselves subject to evolutionary changes. This is possible because the operators are themselves pieces of code, though normally not accessible to the evolutionary process. But meta-level  modification can work, and in fact enables the process to become more efficient \cite{kantschik1999empirical,kantschik1999meta}. The evolutionary process at work here selects improved versions of stochastic operators based on their effect in improving the population fitness. The outcome of self-modification of code is generally not predictable, and it is the post hoc fitness measurement after execution of operators that manipulate the code that allows evolution to produce better operators. Another example of the same principle is at work in self-modifying Cartesian Genetic Programming \cite{harding2010developments}. Here, operators exist that can determine how many inputs a system can access, thus allowing the system itself to determine what dimensionality the input to the programs should have. This allows such systems to solve entirely general classes of problems, as exemplified for the example of a logic problem in \cite{harding2009self}.

\subsubsection{Dynamic categorical frameworks}

{Rosen's Relational Biology \cite{rosen1991life} deployed category 
theory to model the functional organisation of living systems, most notably through his $(M,R)$ metabolism-repair systems, capturing the property of closure to efficient causation (see section~\ref{sec:biology}). 
Rosen's approach foregrounded \emph{relations} and \emph{mappings} over
material substrate, insisting that the organisation of living systems could not be reduced to any mechanism. Rosen argued that living systems are `complex' in a precise technical sense: they admit no complete algorithmic description and cannot be fully captured by any Turing-equivalent simulation \cite{rosen1991life}.}

{Memory Evolutive Systems (MES \cite{ehresmann2007memory}), developed from category theory, addresses a foundational limitation of static categorical models by developing a hierarchical architecture in which new levels of organisation can emerge through `complexification', allowing the system's modelling relations, the metamodels themselves, to evolve. 
MES realises a multivalent logic to capture the context-dependent character of living systems. 
Wandering Logic Intelligence (WLI \cite{simeonov2002}) contributes a non-axiomatic, situation-aware spatio-temporal logic capable of context-dependent reasoning across ambiguous, multi-scale situations.} 

{MES and WLI are integrated in WLIMES \cite{Ehresmann2012,simeonov2017wlimes};
MES provides categorical structure that can evolve at the level of its own organisation, WLI supplies the contextual reasoning required to navigate the full spectrum of behaviours found in living systems. The resulting capacity for second-order logical operations -- reasoning not just within fixed categories but {about} the categorical structures themselves -- represents, in principle, a genuine step beyond first-order approaches and toward frameworks capable of authentic self-modification. Applications cited for WLIMES, in neuroscience \cite{ehresmann2015conciliating} and personalised medicine \cite{simeonov2017wlimes}, remain preliminary.}

{The Integral Biomathics (IB) programme \cite{Simeonov2012,simeonov2013insights,simeonov2015integral} seeks frameworks expressive enough to accommodate self-reference, hierarchical emergence, temporal dynamics, and the evolution of organisational structure itself. IB draws upon Rosen's relational, categorical approach and highlights the WLIMES framework as significant developments.}

{The mathematical frameworks of dynamic categorical hierarchies and second-order
logic build a larger formal container, a more expressive mathematics, one that
can represent self-referential organisation without entering its generative logic. These approaches must, at the point of implementation, resolve certain undecidable aspects \cite{Salthe2013}. The apparent paradoxes of self-reference is the pointed instance of this. MES complexification currently uses externally parameterised dynamics, rather than any kind of endogenous, self-referential time. In this respect, the approach shies away from paradox, remaining outside the constitutive dynamic it describes: the door is reframed but not opened.}

\subsubsection{Reflection and self-modifying models}\label{sec:comp-tools:self-mod}

The Banzhaf et al. \cite{Banzhaf-2016} definition of
open endedness requires that models and metamodels be dynamic, be able to change as the system evolves in time (sec.\ref{sec:open}).  However, conventional mathematical modelling techniques and computer programming languages are static:
they do not change,
and they have no facilities for their models to change,
as the system evolves.

The answer is clear: we need a new generation of dynamical and self-referential programming tools, that can change their own self and their own models through time,
and the accompanying theory to support their development, and support the analysis of the systems they generate.
In sec.\ref{sec:gp}
we see an initial move in this direction:
using an evolutionary algorithm to generate computer programs with properties defined by a fitness function.
The GP system generates novel models (programs) and sometimes metamodels (evolutionary operators), but these are not fully \textit{self}-modifying: they are generated by an external evolutionary algorithm.
In this section, we go a step further,
and discuss the requirements for a fully dynamic system that can self-modify.

A self-modifying computational model is explicitly \textit{self-modifying code}.
Software engineers do not want self-modifying code,
because they cannot control it, predict its behaviour, or statically analyse its properties.
Artificial Life researchers want it, for exactly the same reasons.
Some argue for an intermediate position: allowing restricted self-modification for certain applications  \cite{Christen2022}.

Low-level self modification can happen with assembly language programs.
If the code has access to its own memory space, it can use its write operations to overwrite, and hence modify, the code itself.
(In the early days of computing, when computer memory was scarce, such techniques were used to pack more program into a smaller space. Modern high level languages typically keep their program and data areas segregated to prevent such an occurrence.)
The Artificial Life research domain of automata chemistries depends on self-modifying code
(see, for example, Tierra \cite{Ray1992}, Avida \cite{ofria2004avida}, and Stringmol \cite{SS-ECAL09c}).
Typically, a special-purpose assembly language is defined (although some work uses standard languages \cite{Aguera_y_Arcas2024}), and a simulation with multiple `organisms', each being a small assembly language program, is allowed to evolve. 
{Physis \cite{Egri-Nagy2003} is an early attempt at a self-modifying assembly-level metalanguage (referred to as a self-modifying `genotype-phenotype mapping' in the related literature).}
Stringmol has been shown to exhibit a type-1 and type-2 novelty \cite{Stepney2020:ALife:innov}.

In automata chemistries, the rules for evolutionary operators such as mutation tend to be hard-wired into the system,
so do not themselves evolve.
They are part of the underlying `physics' as opposed to the higher-level `biology' of the simulation, violating the \textit{Everything's Soft} design principle \cite{Hickinbotham-ALJ16}.
The von Neumann self-replication architecture \cite{vonNeumann-1966} is one approach to overcoming this, allowing components to be encoded on a genome, then expressed as active `machines' of the system.
For example, a Stringmol system has been developed conforming to the architecture, and displays a range of innovative behaviours \cite{Clark2017-semclose}.
The EvoMachina \cite{Hoverd-CCS-2016} system is designed on analogous lines,
as a meta-evolutionary algorithm.
It has a Mutator machine that mutates other machines, 
and a MutatorMutator that mutates Mutator machines, and hence can self-mutate.
Such a system would probably never evolve a mechanism such as crossover -- the artificial `physics' is not sufficiently rich -- but a system that had Mutator and Crossover Machines that could self-mutate and self-crossover could potentially evolve a range of novel mechanisms.

It is possible, although more difficult, to evolve high level language programs.
However, a full high-level self-referential, self-modifying computational system would need to employ a form of computational reflection \cite{Maes-1997},
where a program can inspect and change its own contents in a programmed manner.
Interpreted (or just-in-time compiled) languages such as Python incorporate such features, allowing code to be generated dynamically, and hence allowing the model (fig.\ref{fig:metamodel}) to have dynamics.
A proposal for such a system in an open ended simulation setting is outlined in \cite{SS-ECAL11-OEE}.
Developing a method for using such a language  (metamodel) to generate dynamic models (programs) of open ended systems remains an open research topic;
developing a dynamic metalanguage for generating dynamic languages even more so.

\subsection{Tools for analysing self-modifying systems}

\begin{quote}
    \textbf{What are appropriate analysis tools for interrogating self-modifying, evolving, open systems?}
\end{quote}
\noindent
An open ended system will eventually move outside its current model of behaviour, and hence outside any measure based on that model \cite{Stepney+Hickinbotham:2023}.
This makes designing analysis tools challenging.

One set of tools comes from behavioural neuroscience, which is accustomed to wrestling with multiscale problems that span the range of levels from molecular details of synaptic proteins to psychoanalysis \cite{pezzulo2016top}. Thus, the realization that neural computations evolved from much more ancient capacities that all cells possess \cite{fields2020,baluska2016,lyon2006} encourages the porting of conceptual tools and biophysical methods from the behavioural sciences to many novel substrates
\cite{davies2023,stern2020,mcgivern2020}.

Another set of tools could be the use of Genetic Programming (sec.\ref{sec:gp}) to mimic other self-modifying systems. For example, if a GP system can be built that mimics a real economy with its self-modification tendencies built into the GP system, we might be able to learn something about that economy from the algorithmic model the evolutionary process creates in code. 

The domain of Artificial Life includes substantial work on defining, building and analysing open ended systems \cite{Banzhaf-2016,Packard2019,Packard2019ii,Channon2024}.  
Analysis tools exist, for example, the MODES (`Measurements of Open-ended Dynamics in Evolving Systems') toolbox \cite{dolson2019modes}, which has been applied to a range of self-modifying systems, including ecologies and major evolutionary transitions.

\subsection{Applications}
We now turn to some attempts to apply these tools in the foundations of biology and economics.

\subsubsection{Biology: self-repair and self-production}\label{sec:biology}

Two influential approaches in the foundations of biology have been the $(M, R)$--systems (metabolism-repair systems) of Robert Rosen \cite{rosen1991life}, and the concept of autopoiesis of Umberto Maturana and Francisco Varela \cite{maturana1991}.
Both these approaches involve notions of self-reference and self-modification, which they locate as central to the nature of life.

A key idea of Rosen’s is \emph{closure to efficient causation}. 
{Mossio et al \cite{mossio2009computable} express this concept using $\lambda$-calculus (section~\ref{sec:taming}) to show how}
Rosen’s analysis of this leads to the equations
\begin{equation}
    B = f(A), \quad f = \Phi(B), \quad \Phi = \beta(f)
\end{equation}
where $B$ is the result of the material cause $A$ and the efficient cause $f$, $\Phi$ is the `repair’ function,
and the final equation corresponds to `replication’
(see also fig.\ref{fig:repclosure}b). 
Rosen avoids an infinite regress of these causal explanations by taking $\beta = B$.
This sequence of equations regards items such as $B$ as both functions and arguments, just as we saw above. 
It is possible to find a $\lambda$-calculus term providing a uniform solution to these equations \cite{mossio2009computable}. 
This in turn can be interpreted mathematically in terms of domain theory.
Related ways of interpreting $(M, R)$--systems using coalgebras in the guise of non-well-founded sets are described in \cite{chemero2007complexity}.
Korbak \cite{Korbak2023} provides an interpretation that unwinds the circular references through time, in the same way we suggest in fig.\ref{fig:SR-view}b.
Thus we see some lineaments of a rigorous theory underpinning Rosen’s ideas using the mathematics of self-reference.
A remaining challenge is to bridge the (large) gap between these mathematical sketches and plausible biological mechanisms \cite{Hofmeyr2021,Korbak2023}.

We find similar ideas in mathematical analyses of autopoiesis, some by Varela himself \cite{varela1975calculus,varela1978arithmetic}.
A particularly striking example is \cite{soto2011ouroboros}, which uses the {\it ouroboros} form of self-application $xx$ as a central focus of the discussion.\footnote{The ouroboros form typically refers to a cyclical or self-referential structure, inspired by the symbol of the ouroboros, an ancient icon depicting a serpent or dragon eating its own tail.}
As those authors note, it may be more natural to use an approach based on concurrent processes, rather than the more rigid and sequential notion of function application.

\subsubsection{Economic models: self-modification}

Economic systems are highly dynamic on short time scales and evolutionary on longer ones.  
Recio et al. \cite{recio2022dynamics} developed a dynamic model of the economy that is an analogue of the standard Walrasian general equilibrium model; it is able to generate the standard equilibrium solution when unperturbed.  They then inserted an evolutionary mechanism into the dynamic model.  In the model firms are agents that implement production plans taking into account current prices for inputs and output.  Prices are determined so as to equilibrate aggregate supply and demand for each product.  Firms interact with each other and with consumers through market transactions.  Firms are introduced when profits associated with the production of a particular product are above normal, and eliminated when losses are excessive.  The model thus has dynamics at two levels, that of output and prices, and that of system size as measured by the number of firms, with the number of firms fluctuating around an equilibrium level.  

The evolutionary mechanism consists of the occasional introduction of a firm using a new combination of inputs to produce a product not currently present in the system.  This mechanism enables the system to become larger in both the number and variety of firms (i.e. more equations with more variables).  It also becomes different and more complex, as some products (variables) are removed and others are added, and the average production process becomes more complex (the average number of variables per equation increases).   The system also becomes more efficient as measured by the number of consumers that are supported per unit of the primary input (raw material).   The model is thus self-modifying, though to a limited degree.  As an extension, genetic programming was used to evolve several new types of agents from existing ones.  These were composite agents analogous in their structure and behaviour to co-ops, franchises, and firms with branch operations.

\subsubsection{Game theory: changing numbers of players}
One approach that needs to be developed  is game theory in which the number of players not only changes, but changes as a function of the actions of the players. For example, consider the slime mould {\it Physarum polycephalum}, a syncytial organism (a mass of cytoplasm containing multiple nuclei). When it is extending toward a piece of food, one can take a scalpel and separate the leading few millimetres from the rest of the mass. As soon as this is done, the payoff matrix changes because now there are two individuals. The separated smaller one now faces a choice: exploit the food on its own (and not share it with the large mass, for higher nutrient density) or move backward, and rejoin with the mass first. Interestingly, if it were to rejoin, it would cease to exist, in the sense that the payoff matrix guiding its current decisions would cease to be relevant. 

Depending on what it does, 1 or 2 individuals will result, and the calculus of decision-making will change because its actions not only change its rewards and what happens, but also change the number of individuals that exist and make decisions.  
This ability to change the number of individuals in a biological tissue is not some unique quirk of slime mould physiology: it is what happens in embryogenesis (determining whether singletons, twins, triplets, or just a collection of cells results) and in cancer (where the informational network of cells can be interrupted, leading to cells whose boundary between self and world shrinks; they become amoebas and treat the rest of the body as the external environment \cite{levin2021bioelectrical}). 
While it is bio-electrically coupled to the rest of the tissue and thus partaking in its pattern memories, a cell cannot easily weigh options of defecting and moving elsewhere (metastasis). However, once that connection is broken, this course of action can be entertained, which might lead to a decision to disconnect further, a feedback loop which increases the agency of the cell at the expense of its membership in the collective intelligence of the network, and thus leads to a shrinking of its cognitive `light cone' down to ancient, unicellular goals of reproduction and maximizing metabolism at the expense of the group's former grandiose goals of organ maintenance.

The well-known spatialised, iterated Prisoner's Dilemma model {can be extended} to one in which individuals can not only Cooperate and Defect, but also Merge and Split \cite{Shreesha2025}. To our knowledge, the mathematics needed to understand the behaviour of such systems that are self-modifying in the sense of adding or removing selves in real time does not yet exist.

\subsubsection{Game theory: changing the rules}

A striking program of work by Vassilakis \cite{vassilakis1991rules,vassilakis1992some} aims at extending existing economic  game theoretic models to encompass `rules for changing the rules’, with the ultimate aim of finding equilibria for such games.
The tools from domain theory used to build models of the $\lambda$-calculus, and more generally to build functorial fixpoints as needed for the semantics of programming languages, are used to build universal games generated by a given game, which result by closing the game under the ability to change the rules.
Moreover, solution concepts, which should be applied uniformly to games, are modelled using the tools developed by programming language semanticists to model polymorphism.

This is a remarkable \textit{tour de force}, which uses much of the machinery in domain theory developed up to that point in building these intricate models of economic behaviour, in which self-reference and self-modification are central.
However, the limiting features of domain theory itself
seem to have led to unresolved obstructions to further progress. In particular, the non-set-theoretic nature of powerdomains means that, in lifting standard solution concepts from game theory to the domain theoretic setting, the desired features are not fully captured.

Once again, as in the case of the biological modelling, a gap remains between the mathematical theory and the concrete models one wishes to capture.

\section{{Discussion, implications, and} conclusions}
\label{sec:conclusions}

The primary themes of this paper are self-reference, self-modification, and time.  They seem to be fundamental to an understanding of living systems, but in various ways they are problematic in the context of conventional science.  
{We have shown that progress is being made on these problems, and that there are promising pathways for future substantial progress.  The discussion in this section of remaining problems represents an update of the open questions the paper addresses.}

All living systems, from the simplest organism to human systems, as well as  their meta-systems like science and technology, involve self-reference.  In a sense these systems are defined by their self-reference, whether it is in the genome, a set of laws, or cultural norms.  As Rosen \cite{RosenEssays} said, living systems are systems that model themselves.  
Rosen believed that life is not a special case of material systems, it is the most general case, with physics being the special case \cite[p.13]{rosen1991life}.  He conjectured that life involved at least one currently unknown fundamental principle analogous to those known to physics. 

One candidate for such a principle is \textit{representation}, with representation  fundamental to living systems.  But if we take representation as a fundamental principle, we need to understand it as such, rather than simply invoke it as necessary or convenient.  In biology this problem is likely to become one of reductionism versus emergence as more phenomena come to be understood at the molecular level:  at the level of molecules, representation disappears;  how does it emerge?  This is another instance of the coarse-graining problem mentioned previously.

{We have made explicit use of the concept of representation in the case of representational time.
The concept of a \textit{self} is also an instance of representation.  Indeed, the concept of self-reference itself implies a representation of some kind.  However, }
it may be difficult to generalise representation in a formal way, since it covers such a wide range of phenomena.  For example, many discussions start with a distinction between a representation and the thing represented.  But as we noted in section~\ref{sec:TLR},  in biological morphogenesis the computational medium and its work product are the same; thus meanings are constantly changing because the machine itself is constantly changing.  A similar problem exists with respect to understanding representation within a single cell: the representations of self and environment are implicit in the molecular machinery.  
In this case it may seem that if the representation and the thing represented are identical, there is no representation, but the biological notion of self described in section~\ref{sec:establishing} provides an escape.  If an entity is actively involved in its own becoming, then it can be described as a `self':  
selfhood in living systems implies both purpose and process, and thus an implicit separation of representation and the thing represented that is not present in non-living systems.  
This view has much in common with enactivist stances;
as case studies including these concepts become more common, 
a better understanding of how these various stances interact and complement should emerge.  

{One interesting recent approach to formalising such a representation of time is the work of Bennett \cite{Bennett-thesis,Bennett-mind-smeared}.
He suggests a functional criterion for an organism having a representation of time: (a) the system maintains internal variables that encode path-dependent constraints; (b) these are used to select among possible future trajectories; and (c) the variables may be updated by the experience of the individual or by the evolutionary process to which the population is subjected.  The formulation suggests that representation is in general a multi-level process, i.e. one involving multiple layers of abstraction as formulated in his stack theory \cite{Bennett-thesis}.   This suggests one path to a more formal approach to representational time \cite{Bennett-mind-smeared}.}

Self-reference became problematic when it appeared in formal representations  because of the long history of the problem of paradox in conventional logic.   
Coalgebra offers one promising approach to dealing with systems characterised by self-reference.  It has the advantage of providing the same clarity and certainty offered by earlier mathematics.  However, it is not clear how useful it will be in describing systems as they undergo structural evolution.   

Another solution to the self-reference paradox problem is to include natural time in our treatment of the system.  Introducing time into scientific representations resolves paradox into process, into change.  Introducing time in an appropriate way, however, is not always straightforward.  Many of the major problems in the biological and social sciences involve structural evolution, the transformation of a system into another, different system.  Simulation modelling is the most obvious approach to use since it is inherently temporal, but so far it has mostly been employed in very conservative ways.  Applications to structural evolution, whether in biological or social sciences, have been relatively few, and the degree of evolution modelled relatively limited.  
Some of the more powerful techniques, like genetic programming, were designed to solve optimisation problems, and are almost always used in that way;
however, they can also be used to model open ended evolution, but only up to a point.  As discussed in section~\ref{sec:comp-tools:self-mod}, we still have no practical approaches to modelling changes to the rules by which the system structure evolves, much less ways to model even higher level changes.  

{
The autonomy perspective developed through sections \ref{sec:self} and \ref{sec:tools} places the principle of representation in productive tension. In organisational closure, a system constitutes its own identity through recursive self-production, preceding and conditioning any correspondence relation we might subsequently observe from outside. As Varela states, what we call a representation is not a correspondence given an external state of affairs, but a consistency with a system's own ongoing maintenance of identity \cite[p.~xxii]{varela1979}, descriptions made from a vantage point that is not in the system's operation. Held together, representation as explanatory strategy and autonomy as constitutive process generate precisely the kind of apparent contradiction that, in
Varela's own terms, need not be dissolved: confronting it rather than resolving it may be what opens the wider domain. The question we leave genuinely open is whether representation can serve as a foundational principle, or whether it is itself founded on the autonomous organisation that makes representation possible at all.
}

The most immediate impediment to progress may be cultural: a reluctance to build models that create themselves, transforming autonomously, rather than becoming what the programmer wants them to be.  But the whole point of an evolutionary model is to reveal what the model becomes by itself; that is how we will come to understand the evolutionary possibilities of the system being modelled.  
A longer term impediment is potentially that raised by the issues of second order cybernetics \cite{Foerster1976}.
This was a shift from focussing on the observed systems to (the nature of) observing, along with the centrality of recurrence. 
It co-evolved with radical constructivism, as a new kind of epistemology, 
where we construct (or compute) our reality. 
These are deeply challenging notions, 
whose implications require a revisitation of science as a
more reflexive endeavour (second-order science), 
where we as scientists are in the systems we are observing, 
and we must bear the resulting responsibilities.

The most formidable problems, however, are methodological.  As already mentioned we currently have no way of modelling higher order evolutionary changes, like changes to the rules that generate structural change in the system.  Without such hierarchical evolutionary modelling techniques it will be impossible to model the transformations that characterise social systems.  There are many other major methodological problems.  For example, to use most models of real systems for prediction, they must be calibrated.  How do we calibrate an evolutionary model with multiple possible trajectories when we may have many fewer data sets than the model has possible outcomes?  If we calibrate so that the model produces only observed outcomes, then if the model is structurally correct we will have seriously mis-calibrated it, so that the model cannot predict outcomes that could have happened but did not; on the other hand, perhaps the model is structurally incorrect, and those unobserved outcomes could not in fact have happened.  We have no way of  knowing which of these possibilities is the case.  This is a problem that already arises in dynamical models with bifurcating trajectories \cite{DBrown} and at present there is no known method for resolving it.   Because both the predictions and the tests become ambiguous, the problem also triggers a cultural one by undermining a fundamental belief of the scientific community: that progress depends on making predictions and testing them.  

There are many such methodological problems, and we may expect that more will arise.  They are frustrating, but they are a sign that we are making progress, that we are addressing new kinds of problems.  Many of them will arise in specific disciplines and be unimportant in others, so the methodological future is likely to be complex and messy.  The solutions will arise in an evolutionary way: we will generate and test many models across many fields, and gradually we will evolve an ecosystem of models that are more or less successful as well as more or less supportive or at least consistent with each other.

\section*{Acknowledgements}

\appendix

\paragraph{Ethics.}{No ethical approvals were required for this work.}


\paragraph{Author Contributions.}{Author order is alphabetical.
All authors contributed to the workshop, and to writing sections of the text
from their diverse philosophical stances.
W.B., S.S., and R.W. took overall conceptual and editing responsibility.
All authors approved the final version.}


\paragraph{Funding.}{
We gratefully acknowledge funding from Memorial University of Newfoundland and Labrador, under the University Research Professorship program to W.B., for making possible the 2023 workshop on ``Time, Life and Self-reference''.
M.L. gratefully acknowledges the support the John Templeton Foundation (via grant 62212), the Templeton World Charity Foundation (via grant TWCF0606), and of Karen Fries.
}

\paragraph{Acknowledgements.}{Our thanks to the contributions of the other workshop attendees, Albrecht von Müller and David Thompson.
S.S. thanks A.J. López-Díaz for bringing \cite{Korbak2023} to her attention.
}

\paragraph{Disclaimer.}{The opinions expressed in this publication are those of the author(s) and do not necessarily reflect the views of the John Templeton Foundation or the Templeton World Charity Foundation.}

\bibliographystyle{plainnat}
\setlength{\bibsep}{2pt plus 0.3ex}
\bibliography{TLS}

@article{kim2006,
author={Kim, Jaegwon},
title={Emergence: Core ideas and issues},
journal={Synthese}, 
volume={151},
issue={3},
year={2006},
pages={547-559}
}

@article{recio2022dynamics,
  title={From Dynamics to Novelty: An Agent-Based model of the economic system},
  author={Recio, Gustavo and Banzhaf, Wolfgang and White, Roger},
  journal={Artificial Life},
  volume={28},
  number={1},
  pages={58--95},
  year={2022},
}

@incollection{mead1968,
  title={The Cybernetics of Cybernetics},
  author={Mead, Margeret},
  booktitle={Purposive Systems},
  editor={Heinz von Foerster and John D. White and Larry J. Peterson and John K. Russell},
  pages={1--11},
  year={1968},
  publisher={Spartan Books}
}

@article{glanville2013,
  title={Radical Constructivism = Second Order Cybernetics},
  author={Glanville, Ranulph},
  journal={Cybernetics and Human Knowing},
  volume={19},
  pages={27--42},
  year={2013},
}

@article{harding2010developments,
  title={Developments in Cartesian Genetic Programming: Self-modifying CGP},
  author={Harding, Simon and Miller, Julian F and Banzhaf, Wolfgang},
  journal={Genetic Programming and Evolvable Machines},
  volume={11},
  pages={397--439},
  year={2010},
  publisher={Springer}
}

@inproceedings{harding2009self,
  title={Self modifying Cartesian Genetic Programming: Parity},
  author={Harding, Simon and Miller, Julian Francis and Banzhaf, Wolfgang},
  booktitle={2009 IEEE Congress on Evolutionary Computation},
  pages={285--292},
  year={2009},
  organization={IEEE}
}

@inproceedings{kantschik1999empirical,
  title={Empirical analysis of different levels of meta-evolution},
  author={Kantschik, Wolfgang and Dittrich, Peter and Brameier, Markus and Banzhaf, Wolfgang},
  booktitle={Proceedings of the 1999 Congress on Evolutionary Computation, CEC99},
  volume={3},
  pages={2086--2093},
  year={1999},
  publisher={IEEE}
}

@inproceedings{kantschik1999meta,
  title={Meta-evolution in graph GP},
  author={Kantschik, Wolfgang and Dittrich, Peter and Brameier, Markus and Banzhaf, Wolfgang},
  booktitle={Genetic Programming: Second European Workshop, EuroGP’99, G{\"o}teborg, Sweden, Proceedings 2},
  pages={15--28},
  year={1999},
  publisher={Springer}
}

@book{darwin1859,
  title={The Origin of Species by Means of Natural Selection},
  author={C. Darwin},
  year={1859},
  publisher={J. Murray, London}
}

@book{koza1992,
  title={Genetic Programming -- On the Evolution of Computer Programs by Means of Natural Selection},
  author={J. Koza},
  year={1992},
  publisher={MIT Press}
}

@book{bergson1911,
  title={Creative Evolution},
  author={H. Bergson},
  year={1911},
  volume={231},
  publisher={University Press of America}
}

@book{yanofsky2013,
  title={The Outer Limits of Reason: What Science, Mathematics, and Logic cannot tell},
  author={N. S. Yanofsky},
  year={2013},
  publisher={MIT Press}
}

@book{rosen2000,
  title={Anticipatory Systems: Philosophical, Mathematical, and Methodological Foundations},
  author={R. Rosen},
  year={2000},
  XXseries={Research International Series on Systems Science and
 	Engineering, Vol. 1},
  publisher={Springer}
}

@InCollection{russell2016,
	author       =	{Irvine, Andrew David and Deutsch, Harry},
	title        =	{{Russell's Paradox}},
	booktitle    =	{The {Stanford} Encyclopedia of Philosophy},
	editor       =	{Edward N. Zalta and Uri Nodelman},
	note =	{\url{https://plato.stanford.edu/archives/win2016/entries/russell-paradox/}},
	year         =	{2016},
	edition      =	{{W}inter 2016},
    publisher    =	{Metaphysics Research Lab, Stanford University}
}

@incollection{shah2021types,
  title={Types and classification of stem cells},
  author={Shah, Aayush A and Khan, Firdos Alam},
  booktitle={Advances in Application of Stem Cells: From Bench to Clinics},
  pages={25--49},
  year={2021},
  publisher={Springer}
}

@incollection{white2020,
  title={Putting Natural Time into Science},
  author={White, R. and Banzhaf, W.},
  booktitle={From Astrophysics to Unconventional Computing: Essays presented to Susan Stepney on the occasion of her 60th birthday},
  editor={A. Adamatzky and 
	V. Kendon},
  pages={1--21},
  year={2020},
  publisher={Springer}
}

@article{arango2011two,
  title={Two levels of metacognition},
  author={Arango-Mu{\~n}oz, Santiago},
  journal={Philosophia},
  volume={39},
  pages={71--82},
  year={2011},
  publisher={Springer}
}

@book{hofstadter2008,
  title={I Am A Strange Loop},
  author={Hofstadter, Douglas},
  year={2008},
  publisher={Basic Books}
}

@book{dennett1996,
  title={Darwin's Dangerous Idea},
  author={Dennett, Daniel C.},
  year={1996},
  publisher={Simon \& Schuster}
}

@article{uzan2011,
  author={JP Uzan},
  title={Varying Constants, Gravitation and Cosmology},
  journal={Living Reviews in Relativity},
  volume={14},
  pages={2},
  year={2011}
}

@article{webb2001,
  author={JK Webb and MT Murphy and VV Flambaum and VA Dzuba and John D Barrow and CW Churchill and JX Prochaska and AM Wolfe},
  title={Further evidence for cosmological evolution of the fine structure constant},
  journal={Phys.Rev.Lett.},
  volume={87},
  pages={091301},
  year={2001}
}

@article{noble2012,
  title={A Theory of Biological Relativity: No Privileged Level of Causation},
  author={Noble, Denis},
  journal={Interface Focus},
  volume={2},
  number={1},
  pages={55--64},
  year={2012}
}

@book{barwise1996vicious,
  title={Vicious circles: On the mathematics of non-wellfounded phenomena},
  author={Barwise, Jon and Moss, Lawrence},
  year={1996},
  publisher={Center for the Study of Language and Information}
}

@inproceedings{huot2023omegapap,
  title={$\omega$PAP spaces: Reasoning denotationally about higher-order, recursive probabilistic and differentiable programs},
  author={Huot, Mathieu and Lew, Alexander K and Mansinghka, Vikash K and Staton, Sam},
  booktitle={38th Annual ACM/IEEE Symposium on Logic in Computer Science (LICS)},
  pages={1--14},
  year={2023},
  organization={IEEE}
}

@inproceedings{heunen2017convenient,
  title={A convenient category for higher-order probability theory},
  author={Heunen, Chris and Kammar, Ohad and Staton, Sam and Yang, Hongseok},
  booktitle={2017 32nd Annual ACM/IEEE Symposium on Logic in Computer Science (LICS)},
  pages={1--12},
  year={2017},
  organization={IEEE}
}

@inproceedings{staton2016semantics,
  title={Semantics for probabilistic programming: Higher-order functions, continuous distributions, and soft constraints},
  author={Staton, Sam and Yang, Hongseok and Wood, Frank and Heunen, Chris and Kammar, Ohad},
  booktitle={Proceedings of the 31st Annual ACM/IEEE Symposium on Logic in Computer Science},
  pages={525--534},
  year={2016}
}

@article{pelayo2014homotopy,
  title={Homotopy type theory and Voevodsky’s univalent foundations},
  author={Pelayo, {\'A}lvaro and Warren, Michael},
  journal={Bulletin of the American Mathematical Society},
  volume={51},
  number={4},
  pages={597--648},
  year={2014}
}

@book{fong2019invitation,
  title={An invitation to applied category theory: Seven sketches in compositionality},
  author={Fong, Brendan and Spivak, David I},
  year={2019},
  publisher={Cambridge University Press}
}

@techreport{vassilakis1991rules,
  title={Rules for changing the rules},
  author={Vassilakis, Spyros},
  year={1990},
  type={Technical Report},
  note ="\url{https://econpapers.repec.org/paper/wpawuwpga/0211006.htm}",
  number={32},
  institution={Stanford Institute for Theoretical Economics},
}

@article{vassilakis1992some,
  title={Some economic applications of Scott domains},
  author={Vassilakis, Spyros},
  journal={Mathematical Social Sciences},
  volume={24},
  number={2-3},
  pages={173--208},
  year={1992},
  publisher={Elsevier}
}

@article{varela1978arithmetic,
  title={The arithmetic of closure},
  author={Varela, Francisco J and Goguen, Joseph A},
  journal={Cybernetics and System},
  volume={8},
  number={3-4},
  pages={291--324},
  year={1978},
  publisher={Taylor \& Francis}
}

@article{varela1975calculus,
  title={A calculus for self-reference},
  author={Varela, Francisco J},
  journal={International Journal of General Systems},
  volume={2},
  pages={5--24},
  year={1975}
}

@article{chemero2007complexity,
  title={Complexity, hypersets, and the ecological perspective on perception-action},
  author={Chemero, Anthony and Turvey, Michael T},
  journal={Biological Theory},
  volume={2},
  pages={23--36},
  year={2007},
  publisher={Springer}
}

@inproceedings{soto2011ouroboros,
  title={Ouroboros avatars: A mathematical exploration of self-reference and metabolic closure},
  author={Soto-Andrade, Jorge and Jaramillo, Sebastian and Guti{\'e}rrez, Claudio and Letelier, Juan-Carlos},
  booktitle={Advances in Artificial Life (ECAL 2011)},
  pages={763--770},
  year={2011},
  organization={MIT Press}
}

@article{mossio2009computable,
  title={A computable expression of closure to efficient causation},
  author={Mossio, Matteo and Longo, Giuseppe and Stewart, John},
  journal={Journal of Theoretical Biology},
  volume={257},
  number={3},
  pages={489--498},
  year={2009},
  publisher={Elsevier}
}

@book{rosen1991life,
  title={Life itself: A comprehensive inquiry into the nature, origin, and fabrication of life},
  author={Rosen, Robert},
  year={1991},
  publisher={Columbia University Press}
}

@article{kozen2017practical,
  title={Practical coinduction},
  author={Kozen, Dexter and Silva, Alexandra},
  journal={Mathematical Structures in Computer Science},
  volume={27},
  number={7},
  pages={1132--1152},
  year={2017},
  publisher={Cambridge University Press}
}

@article{rutten2000universal,
  title={Universal coalgebra: A theory of systems},
  author={Rutten, Jan JMM},
  journal={Theoretical Computer Science},
  volume={249},
  number={1},
  pages={3--80},
  year={2000},
  publisher={Elsevier}
}

@book{jacobs2017introduction,
  title={Introduction to Coalgebra},
  author={Jacobs, Bart},
  year={2017},
  publisher={Cambridge University Press}
}

@incollection{abramsky1994domain,
  title={Domain theory},
  author={Abramsky, Samson and Jung, Achim},
  booktitle = {Handbook of Logic in Computer Science},
  pages = "1-168",
  year={1994},
  publisher={Oxford University Press}
}

@book{gierz2003continuous,
  title={Continuous Lattices and Domains},
  author={Gierz, Gerhard and Hofmann, Karl Heinrich and Keimel, Klaus and Lawson, Jimmie D and Mislove, Michael and Scott, Dana S},
  year={2003},
  publisher={Cambridge University Press}
}

@book{scott1970outline,
  title={Outline of a mathematical theory of computation},
  author={Scott, Dana},
  year={1970},
  publisher={Oxford University Computing Laboratory, Programming Research Group Oxford}
}

@article{levin2021bioelectrical,
  title={Bioelectrical approaches to cancer as a problem of the scaling of the cellular self},
  author={Levin, Michael},
  journal={Progress in Biophysics and Molecular Biology},
  volume={165},
  pages={102--113},
  year={2021},
  publisher={Elsevier}
}

@article{rahwan2019machine,
  title={Machine Behaviour},
  author={Rahwan, Iyad and Cebrian, Manuel and Obradovich, Nick and Bongard, Josh and Bonnefon, Jean-Fran{\c{c}}ois and Breazeal, Cynthia and Crandall, Jacob W and Christakis, Nicholas A and Couzin, Iain D and Jackson, Matthew O and others},
  journal={Nature},
  volume={568},
  number={7753},
  pages={477--486},
  year={2019}
}

@article{pezzulo2016top,
  title={Top-down models in biology: Explanation and control of complex living systems above the molecular level},
  author={Pezzulo, Giovanni and Levin, Michael},
  journal={Journal of The Royal Society Interface},
  volume={13},
  number={124},
  pages={20160555},
  year={2016}
}

@article{mcgivern2020,
  title={Active materials: Minimal models of cognition?},
  author={McGivern, P.},
  journal={Adaptive Behavior},
  volume={28},
  pages={441--451},
  year={2020}
}

@article{davies2023,
  title={Synthetic morphology with agential materials},
  author={Davies, J. and Levin, M.},
  journal={Nature Reviews in Bioengineering},
  volume={1},
  pages={46-59},
  year={2023}
}

@article{stern2020,
  title={Continual Learning of Multiple Memories in Mechanical Networks},
  author={Stern, M. and Pinson, M.B. and Murugan, A.},
  journal={Physical Review X},
  volume={10},
  pages={031044},
  year={2020}
}

@article{fields2020,
  title={Morphological Coordination: A Common Ancestral Function Unifying Neural and Non-Neural Signaling},
  author={Fields, C. and Bischof, J. and Levin, M.},
  journal={Physiology},
  volume={35},
  pages={16-30},
  year={2020}
}

@article{baluska2016,
  title={On Having No Head: Cognition throughout Biological Systems},
  author={Baluska, F.  and Levin, M.},
  journal={Frontiers in Psychology},
  volume={7},
  pages={902},
  year={2016}
}

@article{lyon2006,
  title={The biogenic approach to cognition},
  author={Lyon, P.},
  journal={Cognitive Processing},
  volume={7},
  pages={11-29},
  year={2006}
}

@book{maturana1991,
  title={Autopoiesis and Cognition: The Realization of the Living},
  author={Maturana, Humberto R and Varela, Francisco J},
  year={1991},
  publisher={Springer}
}

@book{varela1979,
  title={Principles of Biological Autonomy - A New Annotated Edition},
  author={Varela, Francisco J},
  publisher={MIT Press, Cambridge, USA},
  year={2025}
}

@article{grim1997,
  title={Fractal images of formal systems},
  author={St. Denis, P. and Grim, P.},
  journal={Journal of Philosophical Logic},
  volume={26},
  pages={181-222},
  year={1997}
}

@article{grim1993,
  title={Self-reference and chaos in fuzzy logic},
  author={Grim, Patrick},
  journal={IEEE Transactions on Fuzzy Systems},
  volume={1},
  number={4},
  pages={237--253},
  year={1993}
}

@article{grim1993self,
  title={Self-reference and paradox in two and three dimensions},
  author={Grim, Patrick and Mar, Gary and Neiger, Matthew and St. Denis, Paul},
  journal={Computers \& Graphics},
  volume={17},
  number={5},
  pages={609--612},
  year={1993}
}

@article{mar1991pattern,
  title={Pattern and chaos: New images in the semantics of paradox},
  author={Mar, Gary and Grim, Patrick},
  journal={Nous},
  volume={25},
  number={5},
  pages={659--693},
  year={1991}
}

@article{levin2023,
  title={Bioelectric networks: The cognitive glue enabling evolutionary scaling from physiology to mind},
  author={Levin, M.},
  journal={Animal Cognition},
  volume={26},
  pages={1865--1891},
  year={2023}
}

@article{levin2023a,
  title={Darwin's agential materials: Evolutionary implications of multiscale competency in developmental biology},
  author={Levin, M.},
  journal={Cell. Mol. Life Sci},
  volume={80},
  pages={142},
  year={2023}
}

@incollection{varela1991,
    author = {Varela, Francisco J.},
    title = {Organism: A Meshwork of Selfless Selves},
    booktitle = {Boston Studies in the Philosophy of Science},
    publisher = {Springer},
    year = {1991}
}

@article{DBrown,
  title={Path Dependence and the Validation of Agent-Based Spatial Models of Land Use},
  author={Brown, D and Page, R and Riolo, M and Rand, W},
  journal={International Journal of Geographical Information Science},
  volume={19},
  number={},
  pages={153--174},
  year={2005},
}

@book{RosenEssays,
  title={Essays on Life Itself},
  author={Rosen, Robert},
  publisher={Columbia University Press},
  year={2000}
}

@article(Banzhaf-2016,
  author = "Wolfgang Banzhaf and Bert Baumgaertner and Guillaume Beslon and René Doursat
            and James A. Foster and Barry McMullin and Vinicius Veloso {de Melo} and Thomas Miconi
            and Lee Spector and Susan Stepney and Roger White",
  title = "Defining and Simulating Open-Ended Novelty: Requirements, Guidelines, and Challenges",
  journal = "Theory in Biosciences",
  volume = 135,
  number = 3,
  pages = "131-161",
  doi = "10.1007/s12064-016-0229-7",
  year = 2016
)

@inproceedings(Stepney:2021:oee4,
  author = "Susan Stepney",
  title = "Modelling and measuring open-endedness",
  booktitle = "OEE4 workshop, at ALife 2021, Prague, Czech Republic  (online)",
  url = "{http://workshops.alife.org/oee4/papers/stepney-oee4-camera-ready.pdf}",
  year = 2021
)

@incollection{Von_Foerster2003-dv,
    author = "von Foerster, Heinz",
    booktitle = "Understanding Understanding: Essays on Cybernetics and Cognition",
    publisher = "Springer",
    year = "2003",
    title = "Objects: Tokens for (eigen-) behaviors",
    pages = "261--271",
}

@BOOK{Suber-1990,
  title       = "The Paradox of Self-Amendment: A study of Law, Logic, Omnipotence, and Change",
  author      = "Peter Suber",
  publisher   = "Peter Lang",
  year        =  "1990"
}

@article{Stepney-IJGS14,
  author = "Susan Stepney",
  title = "Local and global models of physics and computation",
  journal = "International Journal of General Systems",
  volume = 4,
  issue = 7,
  pages = "673-681",
  doi = "10.1080/03081079.2014.920995",
  year = 2014
}

@phdthesis{Bergson-1889,
  author = "Henri Bergson",
  title = "Time and Free Will",
  year = 1889,
}

@book{Proust-recherche-1913,
  author = "Marcel Proust",
  title = "{\`A} la recherche du temps perdu",
  year = "1913--1927", 
  publisher = "Grasset"
}

@book{Hawking-BHoT-1988,
  author = "Stephen W. Hawking",
  title = "A Brief History of Time",
  year = 1988,
  publisher = "Guild"
}

@book{Clark-surfing-2016,
  author = "Andy Clark",
  title = "Surfing Uncertainty: Prediction, Action, and the Embodied Mind",
  year = 2016,
  publisher = "Oxford University Press"
}

@book{Mermin-GR-2005,
  author = "N. David Mermin",
  title = "It's About Time: Understanding Einstein's relativity",
  year = 2005,
  publisher = "Princeton University Press"
}

@book{Carroll-time-2010,
  author = "Sean M. Carroll",
  title = "From Eternity to Here: The quest for the ultimate theory of time",
  year = 2010,
  publisher = "Oneworld"
}

@book{Barbour-time-1999,
  author = "Julian B. Barbour",
  title = "The End of Time",
  publisher = "Phoenix",
  year = 1999
}

@book{Muller-time-2015,
  editor = "Albrecht von M{\"u}ller and Thomas Filk",
  title = "Re-Thinking Time at the Interface of Physics and Philosophy",
  publisher = "Springer",
  year = 2015
}

@book{Muller-now-2016,
  author = "Richard A. Muller",
  title = "Now: The Physics of Time",
  publisher = "Norton",
  year = 2016
}

@incollection(Caves-ch19-2018,
  author = "Leo Caves and Ana Teixeira de Melo and Susan Stepney and Emma Uprichard",
  title = "Time Will Tell: Narrative Expressions of Time in a Complex World",
  chapter = 19,
  pages = "269-284",          
  crossref = "Walsh-2018"
)

@book(Walsh-2018,
  editor = "Richard Walsh and Susan Stepney",
  title = "Narrating Complexity",
  booktitle = "Narrating Complexity",
  publisher = "Springer",
  year = 2018
)

@inproceedings(Stepney:2023,
  author = "Susan Stepney",
  title = "Life as a Cyber-Bio-Physical System",
  booktitle = "Genetic Programming -- Theory and Practice (GPTP XIX), Ann Arbor, MI, USA, June 2022",
  editor="L. Trujillo and S. Winkler and S. Silva and W. Banzhaf ",
  pages = "167-200",
  publisher = "Springer",
  doi = "10.1162/artl_a_00399",
  year = "2023"
)

@book{Carse-1986,
  author = "James P. Carse",
  title = "Finite and Infinite Games",
  publisher = "Ballantine",
  year = 1986
}

@book{Suits-2014,
  author = "Bernard Suits",
  title = "The Grasshopper: Games, Life, and Utopia",
  edition = 3,
  publisher = "Broadview Press",
  year = 2014
}

@book{Kleppe-2003,
  author = "Anneke Kleppe and Jos Warmer and Wim Bast",
  title = "MDA Explained: the Model Driven Architecture: practice and promise",
  publisher = "Addison-Wesley",
  year = 2003
}

@INPROCEEDINGS{Christen2022,
  title     = "{Curb Your Self-Modifying Code*}",
  author    = "Christen, Patrik",
  booktitle = "{2022 IEEE International Conference on Systems, Man, and
               Cybernetics (SMC)}",
  pages     = "2607-2612",
  year      =  2022,
  publisher = "IEEE"
}

@inproceedings{Hoverd-CCS-2016,
  author = "Tim Hoverd and Susan Stepney",
  title = "EvoMachina: A novel evolutionary algorithm inspired by bacterial genome reorganisation",
  pages = "6pp",          
  booktitle = "2nd EvoEvo Workshop, Conference on Complex Systems 2016, Amsterdam, Netherlands",
  year = 2016
}

@INPROCEEDINGS{Ray1992,
  title     = "{An approach to the synthesis of life}",
  booktitle = "{Artificial Life II}",
  author    = "Ray, Thomas S.",
  publisher = "Addison-Wesley",
  pages     = "371-408",
  year      =  1992
}

@article{ofria2004avida,
  title={Avida: A software platform for research in computational evolutionary biology},
  author={Ofria, Charles and Wilke, Claus O},
  journal={Artificial Life},
  volume={10},
  number={2},
  pages={191--229},
  year={2004},
  publisher={MIT Press}
}

@inproceedings{SS-ECAL09c,
  author = "Simon Hickinbotham and Edward Clark and Susan Stepney and Tim Clarke
            and Adam Nellis and Mungo Pay and Peter Young",
  title = "Molecular microprograms",
  pages = "297-304",
  booktitle = "ECAL 2009, Budapest, Hungary",
  series = "LNCS",
  volume = 5777,
  publisher = "Springer",
  year = 2011
}

@ARTICLE{Aguera_y_Arcas2024,
  title         = "{Computational Life: How Well-formed, Self-replicating
                   Programs Emerge from Simple Interaction}",
  author        = "Agüera y Arcas, Blaise and Alakuijala, Jyrki and Evans, James
                   and Laurie, Ben and Mordvintsev, Alexander and Niklasson,
                   Eyvind and Randazzo, Ettore and Versari, Luca",
  journal       = "arXiv:2406.19108 [cs.NE]",
  month         =  jun,
  year          =  2024,
  archivePrefix = "arXiv",
  primaryClass  = "cs.NE",
  eprint        = "2406.19108"
}

@article(Hickinbotham-ALJ16,
  author = "Simon Hickinbotham and Edward Clark and Adam Nellis and Susan Stepney and Tim Clarke and Peter Young",
  title = "Maximising the adjacent possible in automata chemistries",
  journal = "Artificial Life",
  volume = 22,
  number = 1,
  pages = "49-75",
  doi = "10.1162/ARTL_a_00180",
  year = 2016
)

@article(Clark2017-semclose,
  author = "Edward B. Clark and Simon J. Hickinbotham and Susan Stepney",
  title = "Semantic closure demonstrated by the evolution of 
            a universal constructor architecture in an artificial chemistry",
  journal = "Journal of the Royal Society Interface",
  volume = 14,
  pages = "20161033",
  doi = "10.1098/rsif.2016.1033",
  year = 2017
)

@inproceedings(Stepney2020:ALife:innov,
  author = "Susan Stepney and Simon Hickinbotham",
  title = "Innovation, Variation, and Emergence in an Automata Chemistry",  
  pages = "753-760",
  doi = "10.1162/isal_a_00265",
  booktitle = "ALife 2020",
  publisher = "MIT Press",
  year = 2020
)

@book{vonNeumann-1966,
  author = "John {von Neumann} and Arthur W. Burks",
  title = "Theory of self-reproducing automata",
  publisher = "University of Illinois Press",
  year = 1966
}

@inproceedings{SS-ECAL11-OEE,
  author = "Susan Stepney and Tim Hoverd",
  title = "Reflecting on Open-Ended Evolution",
  pages = "781-788",
  booktitle = "ECAL 2011, Paris, France",
  publisher = "MIT Press",
  year = 2011
}

@inproceedings{Maes-1997,
  author = "Patti Maes",
  title = "Concepts and experiments in computational reflection",
  pages = "147-155",
  booktitle = "OOPSLA'87",
  publisher = "ACM Press",
  year = 1997
}

@article(Stepney+Hickinbotham:2023,
  author = "Susan Stepney and Simon Hickinbotham",
  title = "On the Open-endedness of detecting Open-endedness",
  volume = "30", 
  number = 3,
  pages = "390-416",
  year = 2024,
  doi = "10.1162/artl_a_00399",
  journal = "Artificial Life"
)

@book{Aczel-HTT-2013,
  author = "Aczel, P. and Ahrens, B. and Altenkirch, T. and others", 
  year = 2013,
  title = "Homotopy type theory: Univalent foundations of mathematics",
  publisher = "The Univalent Foundations Program Institute for Advanced Study"
}

@ARTICLE{Levin2019,
  title     = "{The computational boundary of a ``self'': Developmental
               bioelectricity drives multicellularity and scale-Free Cognition}",
  author    = "Levin, Michael",
  journal   = "Frontiers in Psychology",
  volume    =  10,
  pages     =  2688,
  year      =  2019,
  doi       = "10.3389/fpsyg.2019.02688",
}

@book{James-1890,
  author = "William James", 
  year = 1890,
  title = "The Principles of Psychology",
  publisher = "H. Holt and Co."
}

@book{Boden-1990,
  author = "Margaret A. Boden", 
  year = 1990,
  title = "The Creative Mind: Myths and mechanisms",
  publisher = "Weidenfeld and Nicolson"
}

@ARTICLE{Blackiston2008,
  title     = "{Retention of memory through metamorphosis: Can a moth remember
               what it learned as a caterpillar?}",
  author    = "Blackiston, Douglas J and Silva Casey, Elena and Weiss, Martha R",
  journal   = "PloS One",
  volume    =  3,
  number    =  3,
  pages     = "e1736",
  year      =  2008,
  doi       = "10.1371/journal.pone.0001736",
}

@ARTICLE{Blackiston2015,
  title     = "{The stability of memories during brain remodeling: A
               perspective}",
  author    = "Blackiston, Douglas J and Shomrat, Tal and Levin, Michael",
  journal   = "Communicative \& Integrative Biology",
  volume    =  8,
  number    =  5,
  pages     = "e1073424",
  year      =  2015,
  doi       = "10.1080/19420889.2015.1073424",
}

@ARTICLE{Shomrat2013,
  title     = "{An automated training paradigm reveals long-term memory in
               planarians and its persistence through head regeneration}",
  author    = "Shomrat, Tal and Levin, Michael",
  journal   = "The Journal of Experimental Biology",
  volume    =  216,
  number    = "Pt 20",
  pages     = "3799-3810",
  year      =  2013,
  doi       = "10.1242/jeb.087809",
}

@INCOLLECTION{Corning1967,
  title     = "{Regeneration and retention of acquired information}",
  author    = "Corning, W. C.",
  booktitle = "{Chemistry of Learning: Invertebrate Research}",
  editor = "W. C. Corning and S. C. Ratner",
  publisher = "Springer",
  pages     = "281-294",
  year      =  1967,
  doi       = "10.1007/978-1-4899-6565-3_18",
}

@ARTICLE{Levin2024,
  title   = "{Self-Improvising Memory: A Perspective on Memories as Agential,
             Dynamically Reinterpreting Cognitive Glue}",
  author  = "Levin, Michael",
  journal = "Entropy",
  volume  =  26,
  number  =  481,
  year    =  2024,
  doi     = "10.3390/e26060481"
}

@ARTICLE{Channon2024,
  title     = "{Editorial introduction to the 2024 special issue on open-ended
               evolution}",
  author    = "Channon, Alastair and Bedau, Mark A and Packard, Norman H and
               Taylor, Tim",
  journal   = "Artificial Life",
  volume    =  30,
  number    =  3,
  pages     = "300-301",
  year      =  2024,
  doi       = "10.1162/artl_e_00445",
}

@ARTICLE{Packard2019,
  title     = "{Open-Ended Evolution and open-endedness: Editorial introduction
               to the open-Ended Evolution I special issue}",
  author    = "Packard, Norman and Bedau, Mark A and Channon, Alastair and
               Ikegami, Takashi and Rasmussen, Steen and Stanley, Kenneth and
               Taylor, Tim",
  journal   = "Artificial Life",
  volume    =  25,
  number    =  1,
  pages     = "1-3",
  year      =  2019,
  doi       = "10.1162/artl_e_00282",
  pmid      =  30933628,
  issn      = "1064-5462,1530-9185",
  language  = "en"
}

@ARTICLE{Packard2019ii,
  title    = "{An Overview of Open-Ended Evolution: Editorial Introduction to
              the Open-Ended Evolution II Special Issue}",
  author   = "Packard, Norman and Bedau, Mark A and Channon, Alastair and
              Ikegami, Takashi and Rasmussen, Steen and Stanley, Kenneth O and
              Taylor, Tim",
  journal  = "Artificial Life",
  volume   =  25,
  number   =  2,
  pages    = "93-103",
  year     =  2019,
  doi      = "10.1162/artl_a_00291",
  pmid     =  31150285,
  issn     = "1064-5462,1530-9185",
  language = "en"
}

@article{dolson2019modes,
  title={The MODES toolbox: Measurements of open-ended dynamics in evolving systems},
  author={Dolson, Emily L and Vostinar, Anya E and Wiser, Michael J and Ofria, Charles},
  journal={Artificial Life},
  volume={25},
  number={1},
  pages={50--73},
  year={2019}
}

@article{pigozzi2025,
  title={Associative conditioning in gene regulatory network models increases integrative causal emergence},
  author={Pigozzi, Federico and Goldstein, Adam and Levin, Michael},
  journal={Nature Communications Biology},
  volume={8},
  number={1},
  pages={1027:1-15},
  year={2025},
  publisher={Nature Publishing Group UK London}
}

@ARTICLE{McMillen2024,
  title     = "{Collective intelligence: A unifying concept for integrating
               biology across scales and substrates}",
  author    = "McMillen, Patrick and Levin, Michael",
  journal   = "Communications Biology",
  volume    =  7,
  number    =  1,
  pages     =  378,
  year      =  2024,
  doi       = "10.1038/s42003-024-06037-4",
}

@ARTICLE{Shreesha2025,
  title     = "{Extending iterated, spatialized prisoners’ dilemma to understand
               multicellularity: Game theory with self-scaling players}",
  author    = "Shreesha, Lakshwin and Pigozzi, Federico and Goldstein, Adam and
               Levin, Michael",
  journal   = "IEEE Transactions on Molecular, Biological, and Multi-scale
               Communications",
  volume    =  11,
  number    =  2,
  pages     = "135-151",
  year      =  2025,
}

@ARTICLE{Goedel1931,
  title     = "{{\"U}ber formal unentscheidbare S{\"a}tze der Principia Mathematica und verwandter Systeme I}",
  author    = "Kurt G{\"o}del",
  journal   = "Monatshefte f{\"u}r Mathematik und Physik",
  volume    =  38,
  pages     = "173-198",
  year      =  1931,
}

@book{Meltzer1962,
  author = "Bernard Meltzer", 
  year = 1990,
  title = "On Formally Undecidable Propositions of Principia Mathematica and Related Systems",
note = "Translation of the German original, Kurt G{\"o}del, 1931",
  publisher = "Basic Books"
}

@book{spencer-brown1969,
  author = {Spencer-Brown, George},
  title = {Laws of Form},
  publisher = {Allen and Unwin},
  year = {1969}
}

@article{kauffman1980,
  author = {Kauffman, Louis H. and Varela, Francisco J.},
  title = {Form dynamics},
  journal = {Journal of Social and Biological Structures},
  volume = {3},
  number = {2},
  pages = {171--206},
  year = {1980}
}

@article{kauffman2005,
  author = {Kauffman, Louis H.},
  title = {EigenForm},
  journal = {Kybernetes},
  volume = {34},
  number = {1/2},
  pages = {129--150},
  year = {2005}
}

@book{ehresmann2007memory,
  title={Memory Evolutive Systems; Hierarchy, Emergence, Cognition},
  author={Ehresmann, Andr{\'e}e C. and Vanbremeersch, Jean-Paul},
  publisher={Elsevier},
  year={2007},
}

@PHDTHESIS{simeonov2002,
  title    = "{The wandering logic intelligence: A hyperactive approach to
              network evolution and its application to adaptive mobile
              multimedia communications}",
  author   = "Simeonov, Plamen L",
  year     =  2002,
  school   = "Technische Universität Ilmenau",
}

@article{ehresmann2012,
  author = "Andrée C. Ehresmann and Plamen L. Simeonov",
  title = "WLIMES, the Wandering LIMES: Towards a Theoretical Framework for Wandering Logic Intelligence Memory Evolutive Systems",
  pages = "105-122",
  crossref = "simeonov2012"
}

@book{simeonov2012,
  title={Integral Biomathics: Tracing the Road to Reality},
  booktitle={Integral Biomathics: Tracing the Road to Reality},
  editor = "Plamen L. Simeonov and Leslie S. Smith and Andrée C. Ehresmann",
  publisher = "Springer",
  year = 2012,
  doi = "10.1007/978-3-642-28111-2"
}

@article{simeonov2013insights,
  title     = {On some recent insights in {Integral Biomathics}},
  author    = {Simeonov, Plamen L. and Gomez-Ramirez, Jaime and Siregar, Pridi},
  journal   = {Progress in Biophysics and Molecular Biology},
  volume    = {113},
  number    = {1},
  pages     = {216--228},
  year      = {2013},
  doi       = {10.1016/j.pbiomolbio.2013.06.001}
}

@article{simeonov2015integral,
  title={Integral biomathics reloaded: 2015},
  author={Simeonov, Plamen L. and Cottam, Ron},
  journal={Progress in Biophysics and Molecular Biology},
  volume={119},
  number={3},
  pages={728--733},
  year={2015},
  doi={10.1016/j.pbiomolbio.2015.10.001}
}

@article{ehresmann2015conciliating,
  title={Conciliating neuroscience and phenomenology via category theory},
  author={Ehresmann, Andr{\'e}e C. and Gomez-Ramirez, Jaime},
  journal={Progress in Biophysics and Molecular Biology},
  volume={119},
  number={3},
  pages={347--359},
  year={2015},
  publisher={Elsevier},
  issn={0079-6107},
  doi={10.1016/j.pbiomolbio.2015.07.004}
}

@article{simeonov2017wlimes,
  title={Towards a first implementation of the {WLIMES} approach in living system studies advancing the diagnostics and therapy in augmented personalized medicine},
  author={Simeonov, Plamen L.},
  journal={Progress in Biophysics and Molecular Biology},
  volume={131},
  pages={349--365},
  year={2017},
  publisher={Elsevier},
  issn={0079-6107}
}

@InCollection{Lycan2023,
	author       =	{Lycan, William},
	title        =	{{Representational Theories of Consciousness}},
	booktitle    =	{The Stanford Encyclopedia of Philosophy},
	editor       =	{Edward N. Zalta and Uri Nodelman},
	howpublished =	{\url{https://plato.stanford.edu/archives/win2023/entries/consciousness-representational/}},
	year         =	{2023},
	edition      =	{{W}inter 2023},
	publisher    =	{Metaphysics Research Lab, Stanford University}
}

@ARTICLE{Gladziejewski2016,
  title     = "{Predictive coding and representationalism}",
  author    = "Gładziejewski, Paweł",
  journal   = "Synthese",
  publisher = "Springer",
  volume    =  193,
  number    =  2,
  pages     = "559-582",
  year      =  2016,
  doi       = "10.1007/s11229-015-0762-9",
}

@ARTICLE{Clark2013,
  title     = "{Whatever next? Predictive brains, situated agents, and the
               future of cognitive science}",
  author    = "Clark, Andy",
  journal   = "The Behavioral and Brain Sciences",
  publisher = "Cambridge University Press",
  volume    =  36,
  number    =  3,
  pages     = "181-204",
  year      =  2013,
  doi       = "10.1017/S0140525X12000477",
}

@book{Pfeifer2006,
  title={How the Body Shapes the Way We Think: A New View of Intelligence},
  author={Rolf Pfeifer and Josh Bongard},
  publisher={MIT Press},
  year={2006},
  doi = "10.7551/mitpress/3585.001.0001",
}

@incollection{SS-Embody,
  author = "Susan Stepney",
  title = "Embodiment",
  chapter = 12,
  pages = "265--288",
  editor = "Darren Flower and Jonathan Timmis",
  booktitle = "In Silico Immunology",
  publisher = "Springer",
  year = 2007
}

@book{Varela1991b,
  title={The Embodied Mind: Cognitive science and human experience},
  author={Francisco J. Varela and Evan Thompson and Eleanor Rosch},
  publisher={MIT Press},
  year={1991},
}

@book{Dennett1987,
  title={The Intentional Stance},
  author={Daniel C. Dennett},
  publisher={MIT Press},
  year={1987},
}

@incollection{Luhmann1986,
  title     = "{The Autopoiesis of Social Systems}",
  author    = "N. Luhmann",
  booktitle   = "Sociocybernetic Paradoxes",
  editor = "F. Geyer and J. {van der Zouwen}",
  publisher = "Sage",
  year      =  1986,
}

@book{Gilbert2010,
  title={Computational Social Science (4 volumes)},
  author={Nigel Gilbert},
  publisher={Sage},
  year={2010},
}

@ARTICLE{Watts2004,
  title     = "{The ``new'' science of networks}",
  author    = "Watts, Duncan J",
  journal   = "Annual Review of Sociology",
  publisher = "Annual Reviews",
  volume    =  30,
  number    =  1,
  pages     = "243-270",
  year      =  2004,
  doi       = "10.1146/annurev.soc.30.020404.104342",
}

@book{Barabasi2016,
  title={Network Science},
  author={Albert-László Barabási and Márton Pósfai},
  publisher={Cambridge University Press},
  year={2016},
}

@ARTICLE{Waddington1942,
  title     = "{Canalization of development and the inheritance of acquired
               characters}",
  author    = "Waddington, C H",
  journal   = "Nature",
  publisher = "Springer",
  volume    =  150,
  number    =  3811,
  pages     = "563-565",
  year      =  1942,
  doi       = "10.1038/150563a0",
}

@book{Foerster1976,
  title={Cybernetics of Cybernetics: Or, the Control of Control and the Communication of Communication},
  editor={Heinz von Foerster},
  year={1976},
  publisher={Future Systems Inc}
}

@ARTICLE{Brand1976,
  title     = "{For God's Sake, Margaret: Conversation with Gregory Bateson and Margaret Mead}",
  author    = "Stewart Brand",
  journal   = "CoEvolutionary Quarterly",
  volume    =  10,
  number    =  21,
  pages     = "32-44",
  year      =  1976,
}

@book{Maturana1987,
  title={The Tree of Knowledge: The biological roots of human understanding: revised edn},
  author={Humberto R. Maturana and Francisco J. Varela},
  year={1987},
  publisher={Shambhala}
}

@ARTICLE{McNeil2004,
  title   = "What's going on with the topology of recursion?",
  author  = "McNeil, Donald H",
  journal = "SEED (Semiotics, Evolution, Energy, and Development)",
  volume  =  4,
  number  =  1,
  pages   = "2--37",
  year    =  2004,
 note = {Available at \url{https://web.archive.org/web/20240206213647/http:/see.library.utoronto.ca/pages/SEED_Journal.html}}
}

@ARTICLE{Korbak2023,
  title     = "{Self-organisation, (M, R)--systems and enactive cognitive
               science}",
  author    = "Korbak, Tomasz",
  journal   = "Adaptive Behavior",
  volume    =  31,
  number    =  1,
  pages     = "35-49",
  year      =  2023,
  doi       = "10.1177/10597123211066155",
}

@ARTICLE{Hofmeyr2021,
  title     = "{A biochemically-realisable relational model of the
               self-manufacturing cell}",
  author    = "Hofmeyr, Jan-Hendrik S",
  journal   = "Biosystems",
  volume    =  207,
  number    =  104463,
  year      =  2021,
  doi       = "10.1016/j.biosystems.2021.104463",
}

@incollection{Longo2021,
	author = {Giuseppe Longo},
	booktitle = {Einstein vs. Bergson: An Enduring Quarrel on Time},
	editor = {Alessandra Campo and Simone Gozzano},
	pages = {375--406},
	publisher = {De Gruyter},
	title = {Today's Ecological Relevance of Bergson-Einstein Debate on Time},
	year = {2021}
}

@phdthesis{bennett-thesis,
  title={How To Build Conscious Machines},
  author={Bennett, Michael T},
  year={2025},
  school={The Australian National University},
  note = {doi:\href{https://doi.org/10.31237/osf.io/wehmg_v1}{10.31237/osf.io/wehmg\_v1}},
}

@article{Bennett-mind-smeared,
  title     = {A Mind Cannot Be Smeared Across Time},
  author    = {Bennett, Michael T},
  year={2026},
  journal   = "arXiv:2601.11620 [cs.AI]",
  doi = "10.48550/arXiv.2601.11620"
}

@ARTICLE{Higgs2015,
  title     = "{The RNA World: molecular cooperation at the origins of life}",
  author    = "Higgs, Paul G and Lehman, Niles",
  journal   = "Nature Reviews. Genetics",
  volume    =  16,
  number    =  1,
  pages     = "7-17",
  year      =  2015,
  doi       = "10.1038/nrg3841",
}

@article{Sole-2025,
  author = "Ricard Solé and Christopher Kempes and Susan Stepney",
  title = "Origins of life: The possible and the actual",
  journal = "Philosophical Transactions of the Royal Society B",
  volume = 380,
  pages = 20240281,
  year = 2025,
  doi = "10.1098/rstb.2024.0281"
}

@ARTICLE{Rice1953,
  title     = "{Classes of recursively enumerable sets and their decision
               problems}",
  author    = "Rice, H G",
  journal   = "Transactions of the American Mathematical Society",
  volume    =  74,
  number    =  2,
  pages     = "358-366",
  year      =  1953,
  doi       = "10.1090/s0002-9947-1953-0053041-6",
}

@ARTICLE{Kolmogorov1968,
  title     = "{Three approaches to the quantitative definition of information}",
  author    = "Kolmogorov, A N",
  journal   = "International Journal of Computer Mathematics",
  volume    =  2,
  number    = "1-4",
  pages     = "157-168",
  year      =  1968,
  doi       = "10.1080/00207166808803030",
}

@incollection{Levin2023-tele,
  title = "Collective Intelligence of Morphogenesis as a Teleonomic Process",
  author = "Michael Levin",
  chapter = 10,
  pages = "175-197",
  booktitle    = "{Evolution ``on purpose'': Teleonomy in living systems}",
  editor   = "Corning, Peter A and Kauffman, Stuart A and Noble, Denis and
              Shapiro, James Alan and Vane-Wright, Richard Irwin",
  year     =  2023,
  publisher = "MIT Press",
  doi = "10.7551/mitpress/14642.003.0013"
}

@ARTICLE{Levin2026,
  title     = "{Machines all the way up and cognition all the way down: Updating
               the machine metaphor in biology}",
  author    = "Levin, Michael and Watson, Richard A",
  journal   = "Seminars in Cell \& Developmental Biology",
  volume    = "177-178",
  number    =  103668,
  pages     =  103668,
  year      =  2026,
  doi       = "10.1016/j.semcdb.2026.103668",
}

@INPROCEEDINGS{Egri-Nagy2003,
  title     = "{Evolvability of the genotype-phenotype relation in populations
               of self-replicating digital organisms in a Tierra-like system}",
  author    = "Egri-Nagy, Attila and Nehaniv, Chrystopher L",
  booktitle = "{ECAL 2003}",
  publisher = "Springer",
  pages     = "238-247",
  series    = "LNCS",
  year      =  2003,
  doi       = "10.1007/978-3-540-39432-7_26",
}

@ARTICLE{Turing1937,
  title   = "{On Computable Numbers, with an Application to the
             Entscheidungsproblem}",
  author  = "Turing, A M",
  journal = "Proceedings of the London Mathematical Society",
  volume  = "s2-42",
  number  =  1,
  pages   = "230–265",
  year    =  1937,
  doi     = "10.1112/plms/s2-42.1.230"
}

@ARTICLE{Friston2010,
  title     = "{The free-energy principle: a unified brain theory?}",
  author    = "Friston, Karl",
  journal   = "Nature Reviews. Neuroscience",
  volume    =  11,
  number    =  2,
  pages     = "127-138",
  year      =  2010,
  doi       = "10.1038/nrn2787",
}

@ARTICLE{Szathmary2015,
  title    = "{Toward major evolutionary transitions theory 2.0}",
  author   = "Szathmáry, Eörs",
  journal  = "PNAS",
  volume   =  112,
  number   =  33,
  pages    = "10104-10111",
  year     =  2015,
  doi      = "10.1073/pnas.1421398112",
}

@book{MaynardSmith1995,
  author = "Maynard Smith, John and Szathmáry, Eörs",
  title = "The Major Transitions in Evolution",
  publisher = "Oxford University Press",
  year = 1995
}

@ARTICLE{Salthe2013,
  title   = "To naturally compute (something like) biology",
  author  = "Salthe, S N",
  journal = "Progress in Biophysics and Molecular Biology",
  volume  =  113,
  number  =  1,
  pages   = "57--59",
  year    =  2013
}
\end{document}